\documentclass{svmult}                     % onecolumn (standard format)
\smartqed  % flush right qed marks, e.g. at end of proof
\usepackage{graphicx,natbib,amsmath}
\usepackage{amsmath,amsfonts,amssymb,bm,url,color}

\newcommand{\EQ}{\begin{equation}}
\newcommand{\EN}{\end{equation}}
\newcommand{\EQA}{\begin{eqnarray}}
\newcommand{\ENA}{\end{eqnarray}}

\newcommand{\Eq}[1]{Eq.~(\ref{#1})}
\newcommand{\Eqs}[2]{Eqs.~(\ref{#1}) and~(\ref{#2})}

\newcommand{\Sec}[1]{Section~\ref{#1}}
\newcommand{\Fig}[1]{Figure~\ref{#1}}

\newcommand{\Figs}[2]{Figs.~\ref{#1} and \ref{#2}}

\def\half{{\textstyle{1\over2}}}

\def\D{{\sc d}{\;}}
\def\L{{\sc l}{\;}}
\def\Dz{{\sc d}}
\def\Lz{{\sc l}}

\newcommand{\qq}{\bm{q}}
\newcommand{\bra}[1]{\langle #1\rangle}
\newcommand{\Gyr}{\,{\rm Gyr}}

\def\ga{\mathrel{\mathchoice {\vcenter{\offinterlineskip\halign{\hfil
$\displaystyle##$\hfil\cr>\cr\sim\cr}}}
{\vcenter{\offinterlineskip\halign{\hfil$\textstyle##$\hfil\cr>\cr\sim\cr}}}
{\vcenter{\offinterlineskip\halign{\hfil$\scriptstyle##$\hfil\cr>\cr\sim\cr}}}
{\vcenter{\offinterlineskip\halign{\hfil$\scriptscriptstyle##$\hfil\cr>\cr\sim\cr}}}}}
\newcommand{\ysci}[5]{: #1, #5, {\em Science }{\bf #2}, #3--#4.}
\newcommand{\ynat}[5]{: #1, #5, {\em Nature }{\bf #2}, #3--#4.}
\newcommand{\ypr}[5]{: #1, #5, {\em Phys.\ Rev. }{\bf #2}, #3--#4.}
\newcommand{\yjgr}[5]{: #1, #5, {\em J.\ Geophys.\ Res. }{\bf #2}, #3--#4.}
\newcommand{\yprl}[5]{: #1, #5, {\em Phys.\ Rev.\ Lett. }{\bf #2}, #3--#4.}
\newcommand{\yptrs}[5]{: #1, #5, {\em Phil. Trans. Roy. Soc. Lond. }{\bf #2}, #3--#4.}
\newcommand{\yprlN}[4]{: #1, #4, {\em Phys.\ Rev.\ Lett. }{\bf #2}, #3.}
\newcommand{\yapjN}[4]{: #1, #4, {\em Astrophys.\ J. }{\bf #2}, #3.}
\newcommand{\yproc}[7]{, ``#4,'' In {\em #5} (ed.\ #6), pp.\ #2--#3.\ #7 (#1).}
\newcommand{\yprocN}[6]{, ``#3,'' In {\em #4} (ed.\ #5), Id.\ #2.\ #6 (#1).}
\newcommand{\yapjlN}[4]{: #1, #4, {\em Astrophys.\ J.\ Lett. }{\bf #2}, #3.}

\newcommand{\ypreN}[4]{: #1, #4, {\em Phys.\ Rev.\ E }{\bf #2}, #3.}
\newcommand{\yrrpN}[4]{: #1, #4, {\em Rep.\ Prog.\ Phys. }{\bf #2}, #3.}

\newcommand{\yab}[5]{: #1, #5, {\em Astrobiol. }{\bf #2}, #3--#4.}
\newcommand{\yabN}[4]{: #1, #4, {\em Astrobiol. }{\bf #2}, #3.}
\newcommand{\ypnas}[5]{: #1, #5, {\em Proc.\ Nat. Acad.\ Soc. }{\bf #2}, #3--#4.}
\newcommand{\yoleb}[5]{: #1, #5, {\em Orig.\ Life Evol.\ Biosph. }{\bf #2}, #3--#4.}
\newcommand{\yol}[5]{: #1, #5, {\em Orig.\ Life }{\bf #2}, #3--#4.}
\newcommand{\yija}[5]{: #1, #5, {\em Int.\ J.\ Astrobiol.\ }{\bf #2}, #3--#4.}
\newcommand{\yijaS}[5]{: #1, #5 {\em Int.\ J.\ Astrobiol.\ }{\bf #2}, #3--#4.}

\newcommand{\yjcpN}[4]{: #1, #4, {\em J.\ Chem.\ Phys. }{\bf #2}, #3.}
\newcommand{\yjour}[6]{: #1, #6, {\em #2} {\bf #3}, #4--#5.}
\newcommand{\yjourN}[5]{: #1, #5, {\em #2} {\bf #3}, #4.}
\newcommand{\ybook}[3]{: #1, {\em #2}, #3.}

\begin{document}

\title{Homochirality: a prerequisite or consequence of life?}

\author{Axel Brandenburg}

\institute{
$^1$Nordita, KTH Royal Institute of Technology and Stockholm University, Stockholm, Sweden\\
$^2$Department of Astronomy, AlbaNova University Center, Stockholm, Sweden\\
$^3$McWilliams Center for Cosmology, Carnegie Mellon University, Pittsburgh, PA 15213, USA
\\
Correspondence:
Nordita, Roslagstullsbacken 23, SE-10691 Stockholm, Sweden\\
\email{brandenb@nordita.org}, Tel: +46 8 5537 8707, mobile: +46 73 270 4376\\
\url{http://orcid.org/0000-0002-7304-021X}
}

\date{\today}

\maketitle

\abstract{
Many of the building blocks of life such as amino acids and nucleotides
are chiral, i.e., different from their mirror image.
Contemporary life selects and synthesizes only one of two possible
handednesses.
Chiral molecules also tend to be optically active and rotate polarized
light in a left-handed sense for many proteins and in a right-handed
sense for many nucleotides and sugars.
In an abiotic environment, however, there are usually equally many left-
and right-handed molecules, so we talk about a racemic mixture.
If homochirality was a prerequisite of life, there must have been
physical or chemical circumstances that led to the selection of a
certain preference.
Conversely, if it was a consequence of life, we must identify possible
pathways for accomplishing a transition from a racemic to a homochiral
chemistry.
There has been significant progress on both approaches.
One of the four elementary forces of nature, the weak force,
responsible for the decay of free neutrons, for example, does give rise
to a preference for chemical reactions between molecules of a certain
chirality, but the effect is very small compared to that of random
fluctuations that are always present.
On the other hand, amino acids from certain meteorites suggest a
preference for the left-handed amino acids -- at least for some of them
-- although the question of contamination is not fully ruled out.
This could be explained by
polarized light, which can give rise to a selection of a net handedness
of biomolecules, but the effect is again small.
Depending on the mechanism that is responsible for generating polarized
light, either both signs or only one sign of polarization are possible.
After a detailed discussion of these ideas and the observational evidence,
we also review alternative ideas where homochirality of any handedness
could emerge as a consequence of the first polymerization events of
nucleotides in an emerging RNA world.
In those proposals, autocatalysis was thought to be a crucial ingredient,
but recent studies show that this is not necessarily the case.
Also, the effect of enantiomeric cross inhibition, the termination
of polymerization with monomers of the opposite chirality, may not be
detrimental, as was originally thought, but it may instead be beneficial.
There are indeed mechanisms that can produce full homochirality through
the combination of both autocatalysis and enantiomeric cross inhibition.
These mechanisms are not limited to nucleotides, but can also occur for
peptides, as a precursor to the RNA world.
The question of homochirality is, in this sense, intimately tied to the
origin of life.
Future Mars missions may be able to detect biomolecules of extant or
extinct life.
We will therefore also discuss possible experimental setups for
determining the chirality of primitive life forms in situ on Mars.
}


\section{Introduction}

The occurrence of handedness in biology is not uncommon.
The difference between our left and right hands is the most obvious
occurrence in the macroscopic world.
In ancient Greek, the word $\chi\epsilon\iota\rho$ means hand, which
explains the origin of the word chirality.
Also some trees exhibit a preference for a left-handed swirl and others
for a right-handed swirl.
Snails are another such example.
In the microscopic world, the biological significance of a preferred
handedness was discovered by \cite{Pasteur1853} by analyzing the
effect of tartaric acid on polarized light.

Polarization is a property of transversal waves when the wave shows
oscillations perpendicular to the direction of propagation.
This is different for sound waves that are longitudinal.
Unpolarized light consists of a superposition of waves with all the
different polarization orientations of the wave plane.
Using a polarizer, which is an optical filter with maximum transmission
for a particular wave plane, one can determine the orientation of
polarization.
It turns out that natural sugar in solution has the property of rotating
the plane of polarization of polarized light in the right-handed sense,
so they are called dextrorotatory, denoted by $(+)$, while
many amino acids in solution rotate polarized light in the left-handed
sense, denoted by $(-)$.
\cite{Pasteur1853} also inspected the shapes of crystals of tartaric
acid under the microscope and found upon separating them that the two
rotate polarized light in opposite senses.

\begin{figure}[t!]\begin{center}
\includegraphics[width=.6\textwidth]{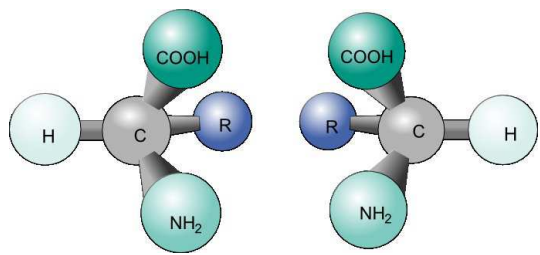}
\end{center}\caption[]{
An amino acid that is chiral whenever the residue R is different from H.
For example, when ${\rm R}={\rm CH}_3$, we have alanine, but when
${\rm R}={\rm H}$, the molecule is glycine, which is the same as its
mirror image, i.e., it is achiral.
(Source: \url{https://chem.libretexts.org/@api/deki/files/19089/molecule.png?revision=1})
}\label{molecule}\end{figure}

\begin{figure}[t!]\begin{center}
\includegraphics[width=\textwidth]{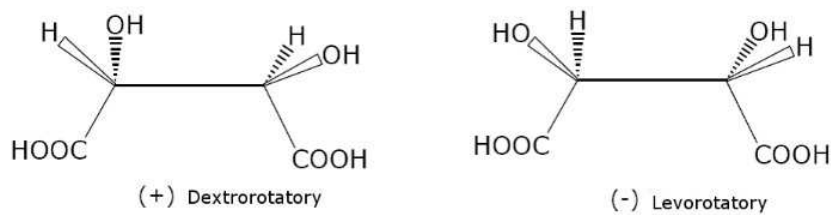}
\end{center}\caption[]{
Dextrorotatory (left) and levorotatory (right) tartaric acid.
Adapted from \cite{Sevin}.
}\label{Sevin1}\end{figure}

Handedness of biomolecules is primarily a consequence of the tetrahedral
shape of the of carbon compounds; see \Fig{molecule}.
If each of the four bonds of the carbon atom connect to a different group,
its three dimensional structure would be different from that of its
mirror image.
In the case of complex molecules, there can be several carbon atoms that
cause a violation of mirror symmetry.
Those carbon atoms are then called chiral centers.
In the case of tartaric acid (\Fig{Sevin1}), there are two chiral centers.
There is then also the possibility that only one of the two chiral centers
is different.
That version is called meso-tartaric acid and it is achiral,
i.e., it is mirrorsymmetric.

\begin{figure}[t!]\begin{center}
\includegraphics[width=\textwidth]{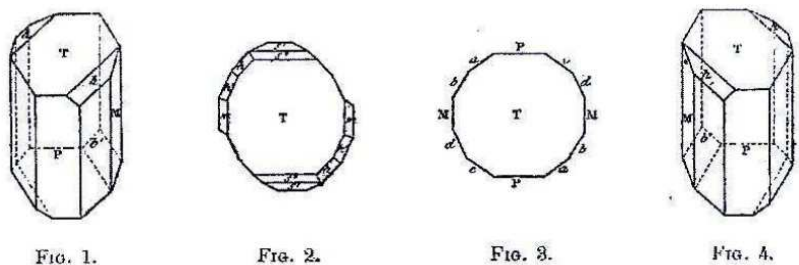}
\end{center}\caption[]{
Original drawing from Pasteur's publication \citep{Masson}
showing dextrorotatory (left, denoted Fig.~1) and
levorotatory (right, denoted Fig.~4) tartaric acid.
Adapted from \cite{Sevin}.
}\label{Sevin2}\end{figure}

There is no immediate connection between the handedness of molecules
(\Fig{Sevin1}) and the handedness hidden in the structure of a crystal
(\Fig{Sevin2}).
In fact, the association of a given chiral structure with left or right
relies on some convention.
This also explains that there is nothing strange in having right-handed
sugars in our DNA and left-handed amino acids in our proteins.
Nevertheless, the very fact that something can occur in two possible
forms that are mirror images of one another is non-trivial and requires
some underlying structure that can also be subdivided into two opposite
mirror images of each other.
It is therefore plausible that one of the two handednesses of the
{\sc l}-tartaric acid molecule crystallizes into macroscopic structures
of one form, and the {\sc d}-tartaric acid into its mirror image
\citep{Der08}.
In the molecular context, these two forms are called enantiomers.\footnote{
We must emphasize that the terminology in terms of levorotatory and dextrorotatory is
quite different from that in terms of \D and \Lz.
Levorotatory/dextrorotatory is the physical property for a compound to induce
the rotation of polarized light to the left/right.
This property is abbreviated $(-)$/$(+)$.
By contrast, \Lz/\D refers to a structural property of a molecule to
denote its handedness, that is solely based on conventions.
This convention only applies to specific biomolecules, including
amino acids and sugars.
This convention has been taken in such a way that all biogenic sugars
are \Dz, and all biogenic amino acids are \Lz.
There is yet another terminology in which R and S refer to a structural
property denoting the handedness of a given chiral carbon in a molecule.
This is also based on a convention, which applies to any chiral organic
compound. 
This convention has been taken in such a way that any chiral carbon can
be assigned uniquely an R or S handedness given a precise set of rules
and is thus a drastically different convention from \Lz/\Dz.
For example,
biogenic \Lz-alanine is dextrorotatory $(+)$ and its chiral carbon is of configuration S;
biogenic \Lz-serine is levorotatory $(-)$ and its chiral carbon is of configuration S;
biogenic \Lz-cysteine is dextrorotatory $(+)$ and its chiral carbon is of configuration R.
Regarding sugars, \Dz-glucose is dextrorotatory and \Dz-fructose is levorotatory.
Note also that common table sugar (sucrose, i.e., a \Dz-glucose--\Dz-fructose dimer)
is dextrorotatory.
If one hydrolyzes it, one obtains a 1:1 mixture of
\Dz-glucose:\Lz-fructose, corresponding to a mixture that is levorotatory,
and thus its common name of ``inverted sugar''.
}

Interestingly, already back then, Pasteur made the statement that the
occurrence of handedness is a demarcating property between living and
nonliving matter; see \cite{GK89}.
Particularly important is Pasteur's discovery of 1857 that certain
unidentified microorganisms had a considerable preference consuming
$(+)$-tartaric acid over $(-)$-tartaric acid; see the review articles
by \cite{Gal08} and especially \cite{Sevin}.
The connection between a preferred handedness of biomolecules and
living matter was reinforced in a number of subsequent papers.
The first important one was by \cite{Frank}, who started his paper by
saying ``I am informed by my colleague Professor W.\ Moore that there
is still widely believed to be a problem of explaining the original
asymmetric synthesis giving rise to the general optical activity of the
chemical substances of living matter.''
He then proposed a model, which contained two key ingredients for
producing a systematic handedness: autocatalysis and mutual antagonism.
Autocatalysis means making more of itself.
This is of course a governing principle of biology, but it is meant here
to be used at the molecular level during polymerization,
i.e., when long chains of shorter monomers are being assembled into a
long macromolecule.
When each building block of the polymer has the same chirality, one says
that it is isotactic.
Mutual antagonism, on the other hand, can be interpreted as the tendency
for a monomer of the wrong handedness to spoil the polymerization,
so that the polymer would no longer be isotactic.

The basic principle discovered by Frank has been governing many of the
ideas reflected in subsequent work in the field of homochirality.
One such example was the work of \cite{FC81}, who also proposed a
mathematical model closely related to that of Frank.
However, there are various other clues to the question of homochirality
on Earth.
One is that there is handedness in one of the four basic forces in nature,
the weak force.
We explain the details below, but this discovery implies that certain
properties of a chiral molecule, for example the dissociation energy,
can be different for the two enantiomers.
The energy difference is usually a {\em very} small fraction --
below $10^{-10}$ of the energy of the molecule itself; see \cite{Bon00}
for a review.
Because of the smallness, it is not obvious that this alone can be
responsible for achieving full homochirality.
Thus, it is generally believed that some amplification mechanism is
always needed.

\begin{figure}[t!]\begin{center}
\includegraphics[width=.8\textwidth]{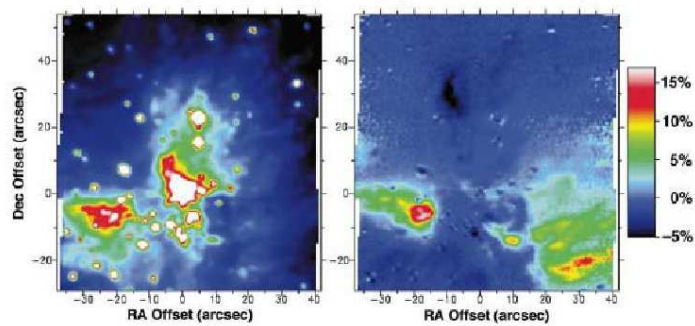}
\end{center}\caption[]{
Circular polarization measurements of the star-forming region
OMC-1 in the Orion constellation.
Note that the circular polarization is predominantly positive
in the bulk of the molecular cloud.
Courtesy of \cite{Bailey98}.
}\label{OMC-1}\end{figure}

An interesting astrobiological connection emerges when considering
circularly polarized light from astrophysical sources.
This is light where the polarization plane rotates with time or position.
Star-forming regions in the Orion constellation have been found to emit
circularly polarized light preferentially in only one of two possible
senses; see \cite{Bailey98,Bailey01}; see \Fig{OMC-1} for an image of
circular polarization in the Orion molecular cloud (OMC).
This is interesting because different enantiomers can dissociate or
degrade differently under the influence of circularly polarized light.
There is further support for this line of thought in that the
chirality of amino acids in space, for example in meteorites,
is found to show a slight preference for the levorotatory ones.

We mentioned already that the connection between chirality and the origin
of life goes back to an early suggestion by Pasteur.
Another connection to astrobiology arises when considering
origins of life on other worlds.
We will return to this at the end when we discuss possible
ways of assessing the reality of extinct or extant life on Mars.

\section{Enantiomeric cross inhibition: the need for homochirality}

We mentioned already that the connection between the origin of
homochirality and the origin of life has been suspected since the early
work of Pasteur.
This connection became more concrete with an important discovery of
\cite{Joyce84}.
He performed experiments with polynucleotide templates, which
facilitate polymerization with the complementary monomers of the same
handedness.\footnote{Instead of polymerization, one sometimes
talks about polycondensation to emphasize the fact that polymerization
implies the removal of water in the reaction.}

\begin{figure}[t!]\begin{center}
\includegraphics[width=\textwidth]{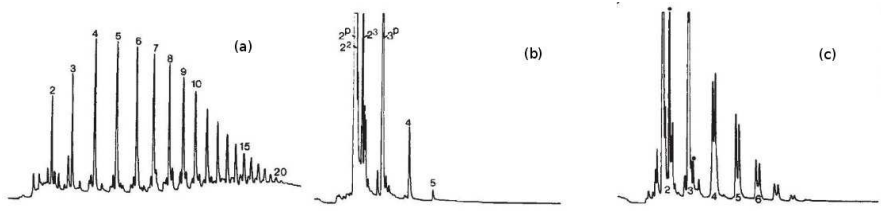}
\end{center}\caption[]{
Chromatograms from the work of \cite{Joyce84} showing template-directed
polycondensation of oligo(G$_{\sc d}$) on poly(C$_{\sc d}$) templates
with \D mononucleotide (left panel), \L mononucleotide (middle panel),
and a racemic mixture of \D and \L mononucleotide (right panel).
}\label{Joyce84cde}\end{figure}

It was thought that polynucleotide templates of one handedness would
direct the pairing with monomers of the same handedness and therefore
favor the selection of nucleotides of the same chirality.
\cite{Joyce84} performed experiments with polymers of dextrorotatory
(\Dz) cytosine (C) nucleobases, poly(C$_{\mbox{\sc d}}$), that are expected
to pair with guanosine (G) mono-nucleotides to form short strands,
oligo(G$_{\mbox{\sc d}}$), along the poly(C$_{\mbox{\sc d}}$).
This was indeed the case and led to the formation of up to 20 base pairs
if the solution contained only monomers that are also dextrorotatory;
see \Fig{Joyce84cde}a.
By contrast, when the solution contained only levorotatory monomers,
no polycondensation occurred; see \Fig{Joyce84cde}b.

This was also expected, because base pairs with opposite handedness do
not fit together.
The surprise came when using a racemic mixture of \D and \L
mono-nucleotides.
A racemic mixture would indeed be expected under prebiotic conditions.
However, in that case there was no significant polycondensation---not
even with the \D mononucleotides; see \Fig{Joyce84cde}c.
Thus, the idea of using template-directed polycondensation to select
only one of two handednesses did not work out.
This phenomenon, which is known as enantiomeric cross inhibition, turned
therefore out to be a major problem for the RNA world \citep{Gil86},
unless there was a reason to expect that only monomers of one handedness
would be around.
\cite{Joyce84} wrote that ``this inhibition raises an important problem
for many theories of the origin of life''.
\cite{Bon91} credited Gol'danskii and Kuz'min in saying ``that a
biogenic scenario for the origin of chiral purity was not viable even in
principle, since without preexisting chiral purity the selfreplication
characteristic of living matter could not occur.''
This is where the discovery of the weak force comes into play.
It provides a reason why one particular handedness might be preferred.
This will be discussed next.

\section{The weak force: non-mirrorsymmetry in nature}

\begin{figure}[t!]\begin{center}
\includegraphics[width=\textwidth]{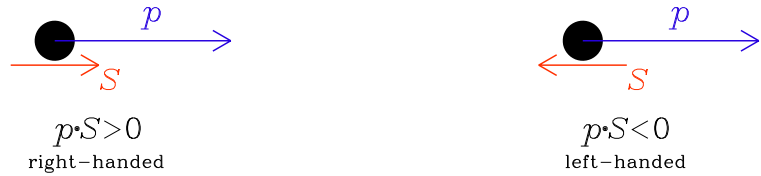}
\end{center}\caption[]{
Illustration of lepton helicity.
The momentum $\bm{p}$ is a polar vector, while the spin $\bm{S}$
is an axial vector, so their dot product is a pseudoscalar,
so it changes its sign when inspected in a mirror.
The electron from the beta decay of a neutron has $\bm{p}\cdot\bm{S}<0$
and is referred to as {\em left-handed}.
}\label{psketch_spin}\end{figure}

At the atomic level, there is the strong and the weak force.
They are two of the four fundamental forces in nature: gravity,
the electromagnetic force, the weak force, and the strong force;
see the early review by \cite{Ulb75} in the astrobiological context.
The weak force is still rather strong compared with gravity ($10^{24}$
times stronger than gravity), but weak compared with the electromagnetic
force ($10^{11}$ times weaker).
The weak force is responsible for the decay of free neutrons, whose
half-time is only about 10 minutes.
The neutron (n) decays then into a proton (p) and an electron (e).
This, as well as the reverse process (electron capture), occur also in the
nuclei of atoms, for example in the decay of radioactive potassium-40
into calcium-40 and argon-40, where the half-time is $1.25\Gyr$.
While there is significant astrobiological significance in this,
for example for dating rocks,\footnote{
Measuring the argon inclusions in solidified rocks is the basis for
determining the age of rocks.
The potassium-40 isotope constitutes only 0.01\% of naturally occurring
potassium. Its half-time is $1.25\Gyr$, making it ideal for geochronology.
The two decay reactions are
\begin{eqnarray}
&&^{40}_{19}{\rm K}\longrightarrow\; ^{40}_{20}{\rm Ca}:\quad
{\rm n}\to\; {\rm p}+{\rm e}+\bar{\nu}_{\rm e}\quad\mbox{($\beta$ decay)},
\\
&&^{40}_{19}{\rm K}\longrightarrow\; ^{40}_{18}{\rm Ar}:\quad
{\rm p}+{\rm e}\to\; {\rm n}+\nu_{\rm e}\quad\mbox{(electron capture)}.
\end{eqnarray}
The latter reaction is responsible for the argon in the atmospheres
of Earth and Mars.
}
we are here concerned with the fact that the electrons from the decay
of neutrons are always left-handed.
This means that the spin of the electron is anti-aligned with its momentum;
see \Fig{psketch_spin}.
At low energies, however, the spin can flip relative to the momentum,
so the handedness of electrons is predominantly a high energy phenomenon.

The fact that electrons produced by $\beta$ decay are chiral is remarkable,
because it means that our physical world is, at least in some respects,
different from its mirror image.
This goes back to a remarkable discovery by \cite{LY56}, which earned them
the Nobel Prize in Physics of 1957 ``for their penetrating investigation
of the so-called parity laws which has led to important discoveries
regarding the elementary particles.''

The connection between the chirality of electrons and that of biomolecules
is not immediately evident.
There are two different ways of establishing a connection between the
handedness imposed by the electroweak force and the handedness in the
biomolecules.
One is through the fact that the bremsstrahlung emission from chiral
electrons rotating around magnetic field lines is circularly polarized with
a sense of polarization that depends on the chirality of the electrons.
This implies that the sense of polarization from bremsstrahlung is
always negative and that this radiation destroys preferentially
right-handed amino acids through photolysis.
This was found by \cite{Goldhaber} and \cite{McVoy} in back-to-back
papers in the Physical Review almost immediately after the influential
paper by \cite{LY56}.
If the idea that circularly polarized light can affect the stability
and selection of biomolecules is to make any sense, one should be able
to discover polarized light in nature.
Interestingly, star-forming regions of OMC-1 in the Orion constellation
have indeed been found to emit right-handed circularly polarized light
\citep{Bailey98}, supporting this basic idea; see \cite{Bailey01} for
a discussion of the astrobiological implications.
However, the circular polarization observed by \cite{Bailey98} occurred
at near-infrared wavelengths and is not related to the mechanism of
\cite{Goldhaber} and \cite{McVoy}, who considered circularly polarized
bremsstrahlung.
\cite{Bailey98} argued that the observed circular polarization is caused
by Mie scattering of unpolarized light, but this mechanism is unrelated
to the weak force.
It is therefore conceivable that also left-handed circularly polarized
light could have been produced in the opposite direction.

Instead of relying on starlight, there is yet another possibility.
Muons, like electrons, belong to the group of fermions that tend to have
a certain handedness.
Muons are about 200 times more massive than electrons and can therefore
be more effective in producing strongly circularly polarized radiation.
Muons occur in the cosmic radiation on Earth.
They are only produced when an energetic cosmic particle hits the Earth's
atmosphere and produces a muon shower.
For this reason, the muons in the cosmic radiation can play a significant
role in affecting the chirality of biomolecules \cite{GB20}.
Unlike the observed circular polarization from the OMC-1 in the Orion
constellation, the sense of circular polarization from this mechanism
is connected with the weak force and therefore, just like in the case
of bremsstrahlung, only one of the two senses are possible, giving
rise to the preferential destruction of right-handed amino acids.

There is another completely different connection between biomolecules
and the weak force.
Quantum-mechanical calculations have shown that the dissociation
energies for \D and \L molecules are slightly different
\citep{Hegstrom,HRS80,MT84}.
Therefore, the \D and \L amino acids in a racemic mixture will degrade
at different rates, which leads to an excess of \L amino acids.

\section{Chiral amino acids in meteorites}

Amino acids have been found in some meteorites \citep{EM97}.
Two particular meteorites are often discussed in connection with
the enantiomeric excess of amino acids: the Murray and the Murchison
meteorites \citep{PC00}.
Those are carbonaceous chondrites, which means that they are
carbon-rich.
They are also rich in organics, as was superficially evidenced by
the smell reported by initial eyewitnesses of the Murchison meteorite.
Interestingly, Table~1.5 of \cite{RGS08} lists 
18 different amino acids that have been found not only
in the Murchison meteorite, but also in the Miller--Urey
experiment \citep{Mil53}.
Twelve of them are not found in proteins on Earth.
This is interesting, because it suggests that those amino acids were
indeed originally present in the meteorite and could not have come from
contamination by life after the meteorite landed on Earth.
Those amino acids that are found on Earth include glycine, alanine,
valine, proline, aspartic acid, and glutamic acid

The sense of the enantiomeric excess is the same in the two meteorites,
corresponding to levorotatory amino acids, but the amount is different
\citep{PC00}.
In addition, there is the possibility that the enantiomeric excess
may be caused by terrestrial contamination \citep{Bada95}.
But, as emphasized above, this would only apply to the six amino acids
that are also found on Earth.
In particular, those amino acids that have the
clearest enantiomeric excess are also those that are most vulnerable
to contamination; see \cite{Ehrenfreund} for a discussion of terrestrial
contaminants in connection with the carbonaceous chondrites
Orgueil and Ivuna.
They are of the type CI (I for Ivuna) and are extremely fragile and
therefore susceptible to terrestrial weathering.
In Orgueil, alanine was found to be racemic and was argued to be
abiotic in origin \citep{Ehrenfreund}.
They could not, however, support the suggestion of terrestrial
contamination with corresponding soil samples.
Incidently, the Orgueil meteorite is also known for a famous contamination
hoax; see \cite{Orgueil64}, who discusses the paper by \cite{Cloez}
claiming the existence of life on the meteoritic parent body a few
weeks after Pasteur's famous lecture to the French Academy on the
spontaneous generation of life.

Among the possible causes for the enantiomeric excess of meteoritic
amino acids, there is the aforementioned effect of circularly polarized
starlight.
Circularly polarized ultraviolet light could have preferentially destroyed
one of the two chiralities through photolysis \citep{ZSS77}.
The experiments of \cite{Bon81} with a \D\L mixture of leucine showed
that right-handed circular polarized light leads to a preferential
destruction of \D leucine, while left-handed circular polarized leads
to a preferential destruction of \L leucine; see also \citep{MT04}
for recent experiments.
To explain the systematic \L excess of amino acids on Earth, one would
need the protosolar nebula to be irradiated by right-handed polarized
light.
Indeed, the star-forming region OMC-1 has been found to emit
right-handed circularly polarized light, supporting this basic idea
\citep{Bailey98,Bailey01,Boyd}.
However, as discussed above, also left-handed circularly polarized
light could have been produced in the opposite direction.
Therefore, any systematic \L excess of amino acids caused by this
mechanism would have been by chance.

The enantiomeric excess found in some amino acids is at most around
1--2\%.
This would be too small to avoid the problem reported by \cite{Joyce84}.
So, even if there is an external effect producing a systematic
enantiomeric excess, we always need an amplification mechanism.
Therefore, we discuss next the Frank mechanism and move then to some
variants of it that avoid either autocatalysis or enantiomeric cross
inhibition.
We begin by explaining first the basic idea.

\section{The basic idea behind the Frank mechanism}

The essence of the mechanism of \cite{Frank} is the {\em combination}
of two ingredients operating in a substrate: catalysis of molecules
for their own production and ``anticatalysis'' that corresponds to some
antagonism or deleterious effect.
He even talks about ``poisoning'' one of the two enantiomers out of
existence.
In fact, he called his simple mathematical model a ``life model'',
suggesting already back then that he was thinking of them as being
processes acting at the moment when the first life emerged.

\begin{figure}[t!]\begin{center}
\includegraphics[width=.6\textwidth]{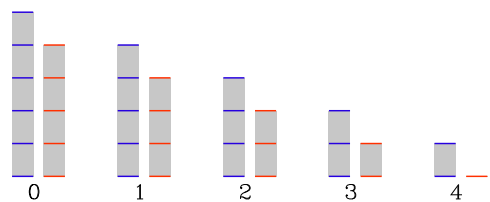}
\end{center}\caption[]{
Sketch showing the effect of enantiomeric cross inhibition only.
Red and blue bars indicate opposite enantiomers, with the blue one
being initially in the majority by one ``unit'', the separation
between subsequent bars.
The gray columns indicate the total amounts, which is 5 units for
the column with blue bars and 4 units for the column with red bars.
In the end, in step 4, only one unit of the enantiomer marked
with blue bars survives.
}\label{psketch_homochiral0}\end{figure}

The essence of Frank's model is perhaps best explained graphically.
For this purpose, it is most instructive to begin with the deleterious
effect by assuming that an equal amount of \D and \L enantiomers eliminate
each other in each reaction step.
This is illustrated in \Fig{psketch_homochiral0}, where we indicate
the amount of \D enantiomers with blue bars and
the amount of \L enantiomers with red bars.

\begin{figure}[t!]\begin{center}
\includegraphics[width=.6\textwidth]{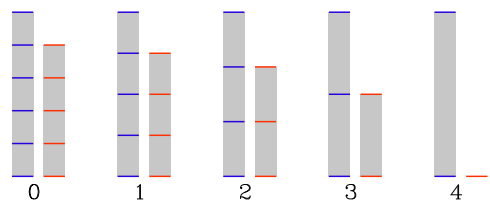}
\end{center}\caption[]{
Similar to \Fig{psketch_homochiral0}, but in each reaction step,
the separations between subsequent bars has been stretched by a certain
factor such that the column with the blue bars retains the same height.
The stretching emulates the effect of autocatalysis.
In the end, again only the enantiomers marked with blue bars survive,
but now, because of the stretching, the amount is no longer small.
}\label{psketch_homochiral1}\end{figure}

We see that, in the end, only \D is left (see the blue bars), but the
amount is very small, namely just as big as the initial difference by
which one of the two enantiomers exceeded the other.
This is why we also need autocatalysis.
Autocatalysis is a process that is not enantioselective, i.e., it works
the same way for the \D and \L enantiomers.
This is demonstrated by stretching out the columns by a factor such
that the highest column always retains the original height; see
\Fig{psketch_homochiral1}.
It is instructive to quantify here the enantiomeric excess (e.e.) as the
ratio of the difference to the sum of the concentrations of right-
and left-handed compounds, i.e.,
\EQ
\mbox{e.e.}=\frac{[D]-[L]}{[D]+[L]}
\label{eedef}
\EN
At each each, the value of $\mbox{e.e.}$
in \Figs{psketch_homochiral0}{psketch_homochiral1} is the same:
$1/(5+4)=1/9$ initially, then $1/(4+3)=1/7$, $1/(3+2)=1/5$, 
$1/(2+1)=1/3$, and finally $1/1=1$.

Frank's paper was mathematical, which may have been a reason why it
was not widely recognized in the biology community at the time.
In fact, it was not quoted in the work of \cite{Joyce84}, who discovered
enantiomeric cross inhibition in the context of nucleotides.
It was only with the paper of \cite{San03} that Frank's mutual antagonism
was identified with enantiomeric cross inhibition.
It was clear from Frank's work that, as long as both reactions,
autocatalysis and antagonism, remain active, the racemic state is
unstable and there will be a bifurcation into a chiral state with an
excess of either \D or \L  enantiomers; see also \cite{San05}, where
enantiomeric cross inhibition was no longer regarded as a problem,
but as an essential ingredient in achieving full homochirality.

In the Frank mechanism, there must be at least a very small initial
imbalance which will then be amplified.
However, this is not a problem because, even if we tried to construct
a purely racemic mixture in the laboratory, there will always remain a
tiny imbalance.
This is just for the same reasons that in a cup of blueberries we will
hardly ever have exactly the same number twice.\footnote{If you take a
cup of blueberries, for example, the exact number varies between 65 and
70 (\url{https://www.howmuchisin.com/produce_converters/blueberries}),
so we must always expect there to be a small imbalance in the number if
we say we have an equal {\em amount} of \D and \L enantiomers.
Mathematically, this imbalance grows with the square root of the number
of molecules (or blueberries) and would be about $\pm10^{12}$ for
one mole with ${\cal N}=6\times10^{23}$ molecules (or $\pm8$ for $65$
blueberries); the fractional imbalance is $1/\sqrt{\cal N}=10^{-12}$
in one mole (or $12\%$ for $65$ blueberries).
}

Given that the racemic state is unstable, the enantiomeric excess,
as defined in \Eq{eedef},
will grow exponentially in time and it does therefore not matter how small
the initial imbalance in the concentrations of \D and \L was.
To demonstrate this more clearly, we use here a figure of \cite{BLL07},
who considered the model of \cite{Plasson}, which we discuss later in
more detail in \Sec{RecyclingFrank}.
This model also has the property that the racemic state is unstable
and that the system evolves toward one of the two homochiral states.

\begin{figure}[t!]\begin{center}
\includegraphics[width=\columnwidth]{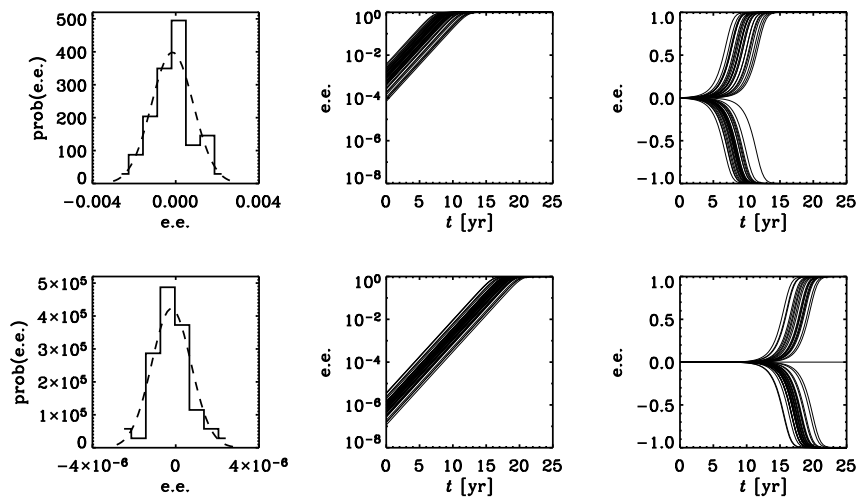}
\end{center}\caption[]{
Probability distribution of the initial enantiomeric excess (e.e.)
for racemic mixtures with $10^6$ and $10^{12}$ molecules together with
the resulting evolution of $\mbox{e.e.}$, both in logarithmic and
linear representations.
The dashed lines give a gaussian fit to the distribution function.
Adapted from \cite{BLL07}.
}\label{pgrowth_array_dist}\end{figure}

In the following, we discuss an ensemble of solutions of the model of
\cite{Plasson} with different realizations or initial states, which
consisted of a racemic mixture of equally many \D and \L enantiomers.
\Fig{pgrowth_array_dist} shows that one always obtains a fully homochiral
state, but in about 50\% of the cases (or in 50\% of the realizations of
the same experiment), one obtains eventually a state with either just
\D enantiomers, and in the other 50\% of the cases or realizations,
one with only \L enantiomers.
When we talk about different cases or realizations, we must realize that
the genesis of life on Earth is just one such realization.
Another one may have occurred on Mars, or in the atmosphere of Venus,
or elsewhere in the Galaxy.
Of course, there is also the possibility of multiple geneses on Earth
alone, with certain lifeforms being either completely or partially
wiped out \citep{DL05}.
The latter case may be particularly interesting in models where we allow
for chemical evolution in models with spatial extent, which will also be
discussed later in \Sec{SpatioTemporal}.

\section{Evidence for autocatalysis}
\label{EvidenceAutocatalysis}

Unlike the process of enantiomeric cross inhibition, where we have
referred to the experiments of \cite{Joyce84}, the actual evidence for
autocatalysis is poor.
In fact, there is only the classical reaction of \cite{Soai} that
exhibits autocatalysis and can lead to a finite enantiomeric excess;
see \cite{Gehring} and \cite{Athavale} for more recent work
clarifying the implications of the Soai reaction.
However, the basic idea of autocatalysis remains plausible, especially
since the discovery by \cite{Altman} and \cite{Cech} that RNA molecules
can exhibit autocatalytic functionality.
This was a very important discovery that earned Sidney Altman and Thomas
R.\ Cech the Nobel Prize in Chemistry in 1989 ``for their discovery of
catalytic properties of RNA.''
It is this mechanism that is at the heart of the idea of an RNA world
\citep{Gil86}.

It is important to realize that the existing evidence for autocatalysis
is irrelevant from an astrobiological viewpoint.
This is particularly clear in view of the fact that the Soai reaction
requires zinc alkoxides as an additional crucial catalyst.
Those compounds are not generally believed to play a role on the early
Earth.

Autocatalysis in the sense of making more of itself is obviously a basic
principle of life, but this is already at a rather complex and not at
the level of individual molecules.
It is therefore possible that autocatalysis does not play a significant role
and that it is rather the process of network catalysis \citep{Plasson15},
i.e., the combined action of different molecules that lead to the desired
appearance of what is in the end equivalent to autocatalysis.
We will return to this in \Sec{RecyclingFrank}, when we discuss a
particular sequence of reactions that, in the end, have the effect of
autocatalysis, even though autocatalysis is not present in any individual
reaction.

\section{The effect of an external chiral influence}

In the beginning of this review, we have discussed extensively the
possibility of a systematic bias resulting eventually from the fact
that the weak force introduces a preference of one of two handednesses
through one or several possible effects.
Those would always favor \L amino acids and \D sugars.
On the other hand, we have now seen that the Frank mechanism can result
in full homochirality of either chirality.
Does this mean that the bias introduced by the weak force is unimportant?
Maybe not quite.
It depends on how strong the external influence is in comparison with
the speed of autocatalysis, which determines the rate of the instability.
This was first discussed in the work of \cite{KN83,KN85} in papers that
appeared just at the time as that of \cite{Joyce84}, but, at the time,
neither of those authors mentioned the work of Frank.

The paper by \cite{KN83,KN85} was in principle quite general and therefore
applicable to other symmetry breaking instabilities.
In essence, the effect of the bias is that it makes the bifurcation
asymmetric.
A symmetric bifurcation is one where the enantiomeric excess (positive or
negative) departs away from strictly zero as some bifurcation parameter
increases.
\cite{San03} identified this bifurcation parameter with the fidelity of
the autocatalytic process, which measures the probability with which
the catalytic process does indeed facilitate the polymerization with
monomers of the same chirality instead of the opposite one.
The fidelity $f$ is unity (zero) when the autocatalytic process always
(never) produces polymerization with the same handedness.

\begin{figure}[t!]\begin{center}
\includegraphics[width=.49\textwidth]{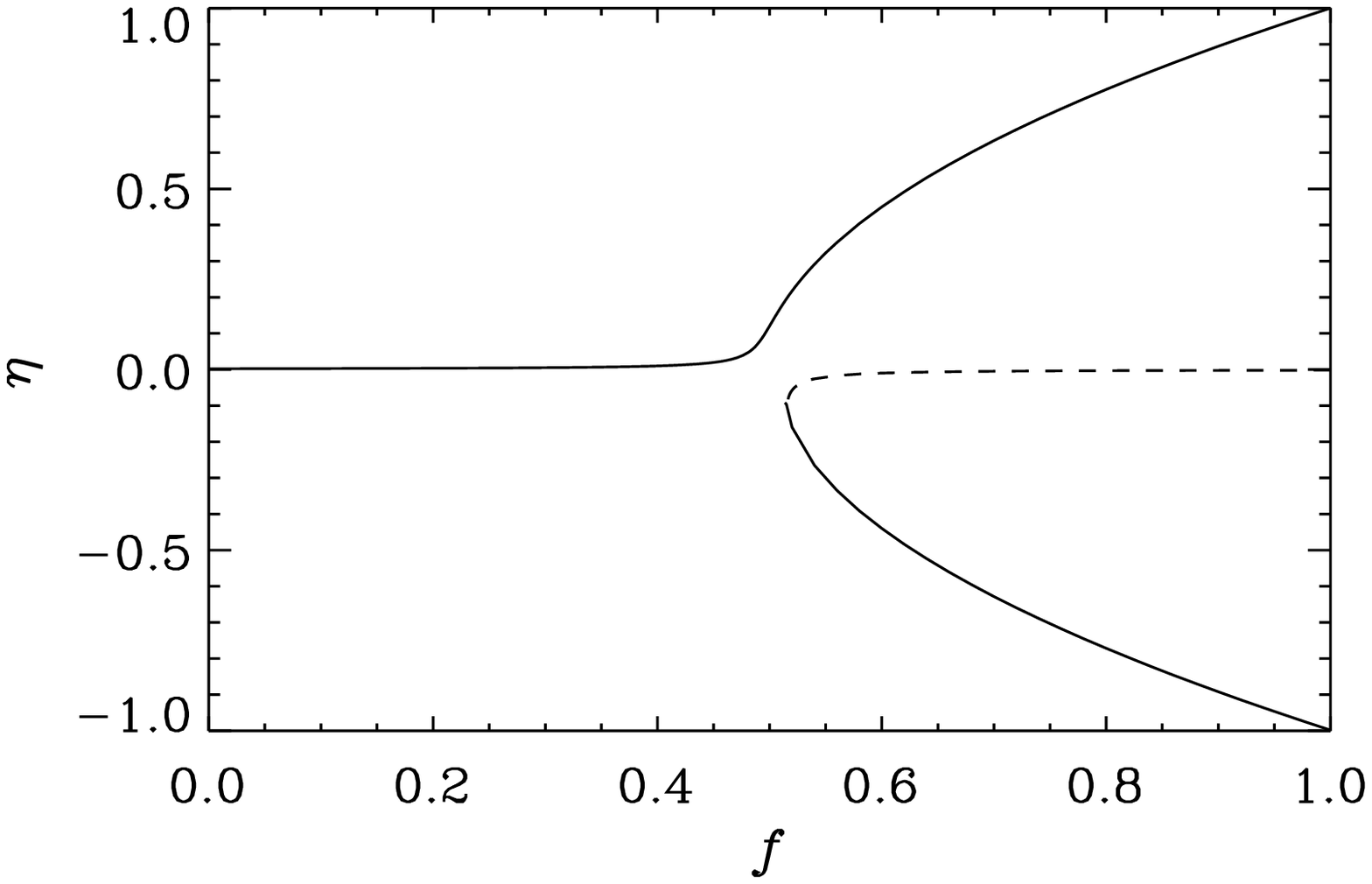}
\includegraphics[width=.49\textwidth]{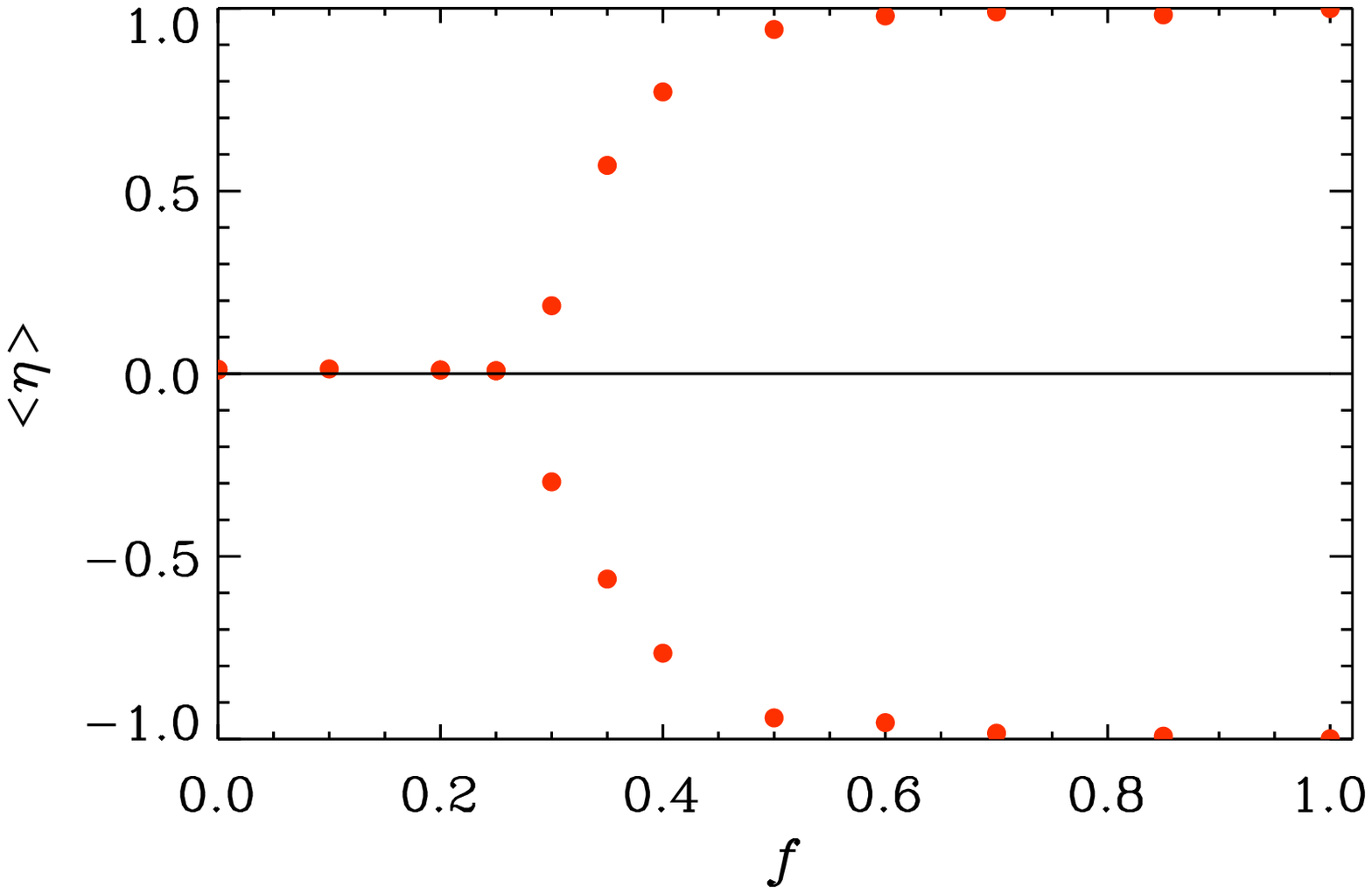}
\end{center}\caption[]{
Bifurcation diagrams showing a slight preference for positive
enantiomeric excess (e.e., here denoted by $\eta$).
The left panel has been adapted from \cite{BAHN}, where
the bifurcation begins for a fidelity $f$ that is clearly below
the otherwise critical value of $f=0.5$.
(The dashed line denotes the unstable solution.)
The right panel has been adapted from \cite{Bra19}, who considered
a stochastic model where $f=0.2$ was assumed.
}\label{pbranches}\end{figure}

In \Fig{pbranches} we show a bifurcation diagram from the work of \cite{BAHN},
where we see that for all values of the fidelity $f$, the solution with
positive enantiomeric excess ($\eta$) is stable.
For $\eta\ga0.5$, the solution with negative $\eta$ is also stable,
but to reach this solution, the initial fluctuations must be large enough.
The complete bifurcation diagram also contains an unstable solution, which
corresponds to the watershed between the two stable branches.
In the left hand plot of \Fig{pbranches}, it is shown as a dashed line.
Similar diagrams have also appeared in the works of \cite{KN83} and later
in the review of \cite{AGK91}.

\section{Polymerization model of Sandars (2003)}

Looking at the chromatographs of \cite{Joyce84}, we see that the ultimate
goal is to assemble long polymers.
For this reason, \cite{San03} developed a polymerization model for
\D and \L nucleotides, where he also allowed for enantiomeric
cross inhibition.
In his model, monomers of the \D and \L forms are being produced at rates,
$Q_D$ and $Q_L$, respectively, that are proportional to same reaction
rate $k_C$ and the concentration of some substrate $[S]$, i.e.,
\begin{equation}
Q_D=k_C[S]\Big\{\half(1+f)C_D+\half(1-f)C_L+C_{0D}\Big\},
\label{QRdef}
\end{equation}
\begin{equation}
Q_L=k_C[S]\Big\{\half(1+f)C_L+\half(1-f)C_D+C_{0L}\Big\},
\label{QLdef}
\end{equation}
where $0\le f\le1$ is the fidelity, $C_L$ and $C_D$ are parameters
describing the global handedness of the system [the concentrations
of the longest possible chains of left- and right-handed polymers for
\cite{San03} and quantities proportional to the masses of all polymers
of the \D and \L forms for \cite{BAHN}].
These parameters are introduced in such a way that for $f>0$ in
\Eqs{QRdef}{QLdef}, $Q_D$ increases with $C_D$, and $Q_L$ increases
with $C_L$.
The parameters $C_{0D}$ and $C_{0L}$ allow for the possibility of
non-catalytic production of left- and right-handed monomers.
They can be different from zero when there is an external bias or
external influence.
When $C_{0D}=C_{0L}=0$, \Eqs{QRdef}{QLdef} show that
$Q_D=k_C[S]C_D$ and $Q_L=k_C[S]C_L$ when $f=1$, while
$Q_D=Q_L=k_C[S](C_D+C_L)/2$ when $f=0$.

\cite{San03} assumed that the catalytic effect depends on the
concentrations of the longest possible chains of left and right handed
polymers.
\cite{BAHN} adopted a similar model, but assumed that $C_D$ and $C_L$
to be proportional to the masses of all polymers of the \D and \L forms,
respectively.
This allowed them to extend the model to much longer polymers without
needing to wait for the longest one to appear before autocatalysis became
possible at all.

\begin{figure}[t!]\begin{center}
\includegraphics[width=.6\textwidth]{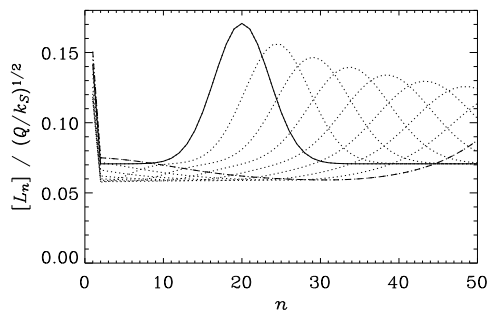}
\end{center}\caption[]{
Normalized concentration $[L_n]$ versus $n$ showing a
wave-like evolution of an initial Gaussian profile (solid line).
All later times are shown as dashed lines, except for the
last time, which is shown as a long-dashed line.
Adapted from \cite{BAHN}.
}\label{pwave}\end{figure}

The full set of reactions included in the model of \cite{San03}
is (for $n\ge2$)
\begin{eqnarray}
L_{n}+L_1&\stackrel{2k_S~}{\longrightarrow}&L_{n+1},
\label{react1}\\
L_{n}+D_1&\stackrel{2k_I~}{\longrightarrow}&L_nD_1,
\label{react2}\\
L_1+L_{n}D_1&\stackrel{k_S~}{\longrightarrow}&L_{n+1}D_1,
\label{react3}\\
D_1+L_{n}D_1&\stackrel{k_I~}{\longrightarrow}&D_1L_nD_1,
\label{react4}
\end{eqnarray}
where $k_S$ and $k_I$ are suitably chosen reaction rates for symmetric
autocatalysis and (non-symmetric) inhibition, respectively.
For all four equations we have the complementary reactions obtained by
exchanging $L\rightleftarrows D$.
The polymerization starts from a large but limited set of monomers that
all begin to develop longer polymers.
Because the number of monomers was limited, the theoretically obtained
chromatograms show a characteristic wave-like motion with increasing time;
see \Fig{pwave}.

\begin{figure}[t!]\begin{center}
\includegraphics[width=\textwidth]{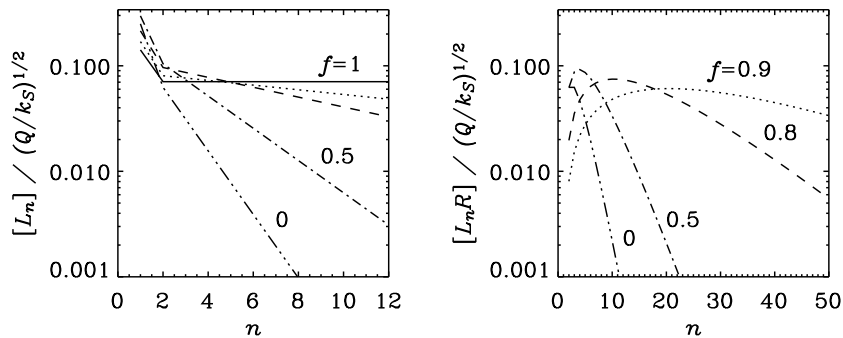}
\end{center}\caption[]{
$[L_n]$ (left) and $[L_nD]$ (right) of equilibrium solutions for different
values of $f$.
For $f=1$ we have $[L_nD]=0$, which cannot be seen in the logarithmic
representation.
Adapted from \cite{BAHN}.
}\label{pspectra}\end{figure}

Particularly important is of course the case where monomers of both
chiralities exist.
In that case, the result depends on the fidelity of the autocatalytic
reactions; see \Fig{pspectra} for such a result.
We see that longer polymers can only be produced when the fidelity is
relatively high.
The lack of sufficient fidelity therefore explains the limited length
of polymers found in the work of \cite{Joyce84}.

\section{Spatiotemporal chirality dynamics}
\label{SpatioTemporal}

In all the chemical reactions discussed so far, the assumption was made
that the system is well mixed.
This means that the concentrations $[D_n]$, $[D_nL]$, etc, are the
same everywhere.
On larger length scales, this assumption must eventually break down.
Even on the scale of alkaline hydrothermal vents, where many scientists
place the origin of life \citep{Rus06} the relevant chemical reactions
would take place within small semiporous cells.
It is then conceivable that similar reactions take place in neighboring
compartments that would be formed by the sulfurous precipitants from
these vents.
\cite{Rus06} draws here an analysis to the chemical gardens that would
allow for a growing arrangement of new compartments, which could act as
primitive cells and would, in principle, allow for Darwinian evolution
as these chemical reactions propagate from one layer of compartments to
the next; see also \cite{Rus14} and \cite{Bar17,Bar19} for more recent
developments.
In each of these compartments, strong spatial gradients and $10^8$-fold
concentration enhancements can be achieved through thermal convective
flows when the aspect ratio of the compartment is sufficiently large
\citep{Braun}.
This setup can also lead to oscillations, which can locally lead to
exponential replication of nuclei acids, analogous to the polymerase
chain reaction \citep{Braun02}.

In the scenario described above, we can no longer talk about a well
mixed system.
Therefore, the concentrations must be regarded as function not only of
time, but also of space.
Because the chemistry in neighboring compartments is loosely coupled by
diffusion terms, there would be spatio-temporal evolution.
In that case, the chemical reaction equations attain a spatial
diffusion term.
The resulting system of equations is usually referred to as
reaction--diffusion equations.
Such models, but for only one instead of several species, have frequently
been employed in modeling the dynamics of diseases such as the black
death \citep{Nob74} or rabies \citep{KAM85,Mur86}.
It has also been used to model the spreading of the novel coronavirus,
where the total number of cases was found to follow a quadratic
or piecewise quadratic growth behavior \citep{Bra20cor}.

\begin{figure}[t!]\begin{center}
\includegraphics[width=.9\columnwidth]{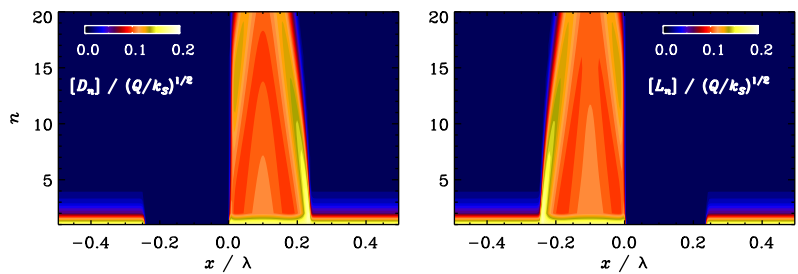}
\end{center}\caption[]{
Color scale plots of $[D_n]$ and $[L_n]$ after $0.8$ diffusion times
as a function of position $x$ and polymer length $n$.
In $0<x/\lambda<0.25$, only \D polymers exist (left) and no \L polymers
at all (right), while in $-0.25<x/\lambda<0$ it is the other way around.
Adapted from \cite{BM04}.
}\label{pchain}\end{figure}

To address the question of homochirality in an extended system,
\cite{BM04} employed a similar approach, but with two or multiple species.
Multiple species occur when we invoke polymers of different length and
composition of different species for the \D and \L forms.
They found that a given species tends to spread through front propagation.
It turned out that, once two populations of opposite chirality
meet, the front can no longer propagate and the evolution comes to a halt.
This result was first obtained in a one-dimensional model, where
the concentrations of {\sc d} and {\sc l} depend on just one spatial
coordinate $x$ and on time $t$.
The result is shown in \Fig{pchain}, for the evolution of short polymers.
These are all regarded as separate species.
The initial condition consists of a small number of monomers of the \L
form at one position (at $x/\lambda=-0.1$ in \Fig{pchain}, where $\lambda$
is the length of the domain) and a three times larger number of monomers
of the \D form at another position (at $x/\lambda=+0.1$).
The theoretical chromatograms are stacked next to each other for each
$x$ position.
We see that in both positions, longer polymers are produced, indicated
by the yellow-reddish colors.
The initially threefold larger number of \D monomers is insignificant,
because the growth is exponentially fast soon saturates at the same
level as that for the \L monomers, when the polymers have reached their
maximum size.
At the same time, polymers of the same handedness can still be produced
by diffusion to the neighboring positions.
This leads to a propagation front.
However, when polymers of opposite chirality emerge at neighboring
positions, the front stops (here at $x\approx0$), while on the other
two sides, the reaction fronts still diffuse further outward.

\begin{figure}[t!]
\includegraphics[width=.41\textwidth]{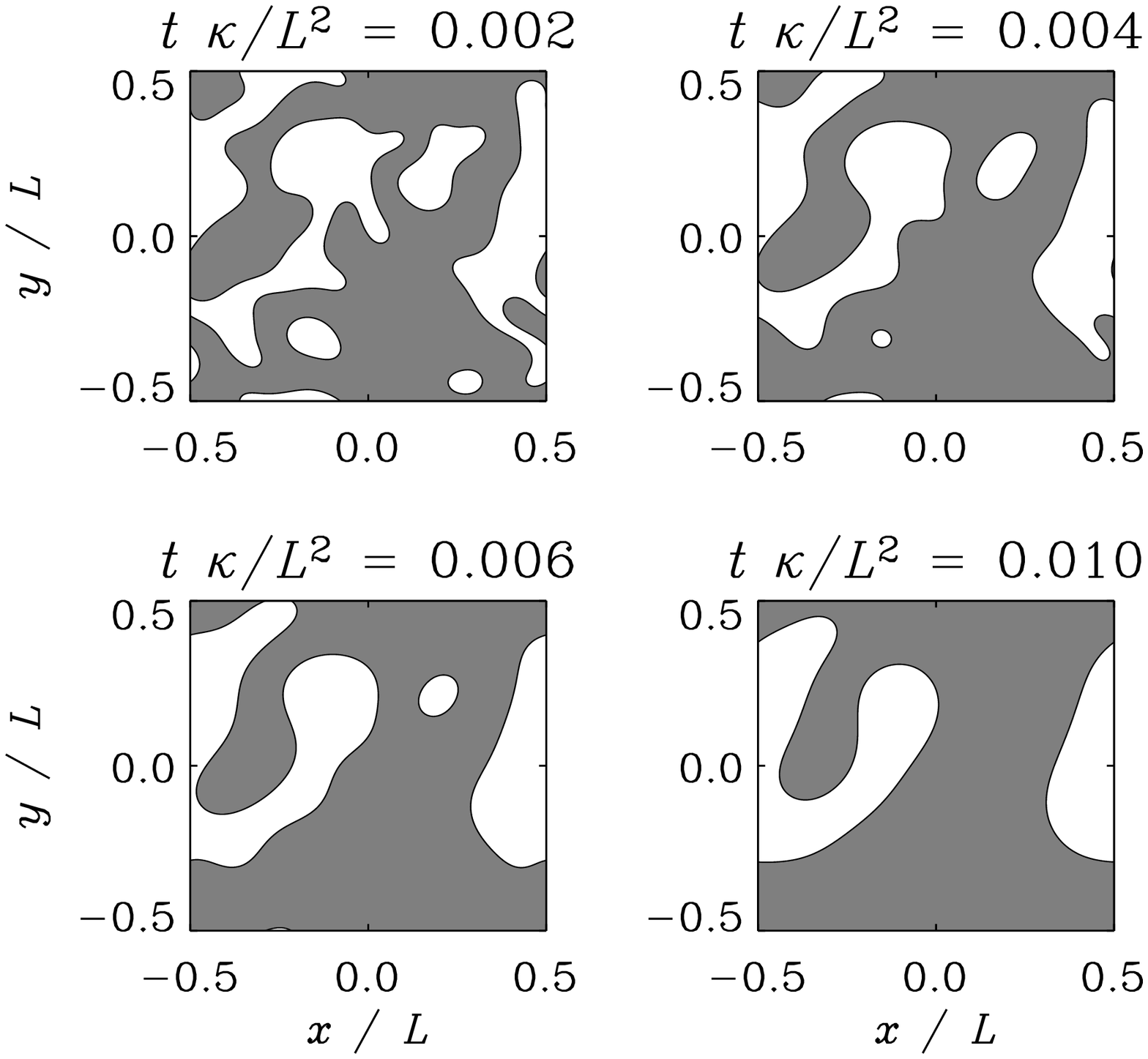}
\includegraphics[width=.58\textwidth]{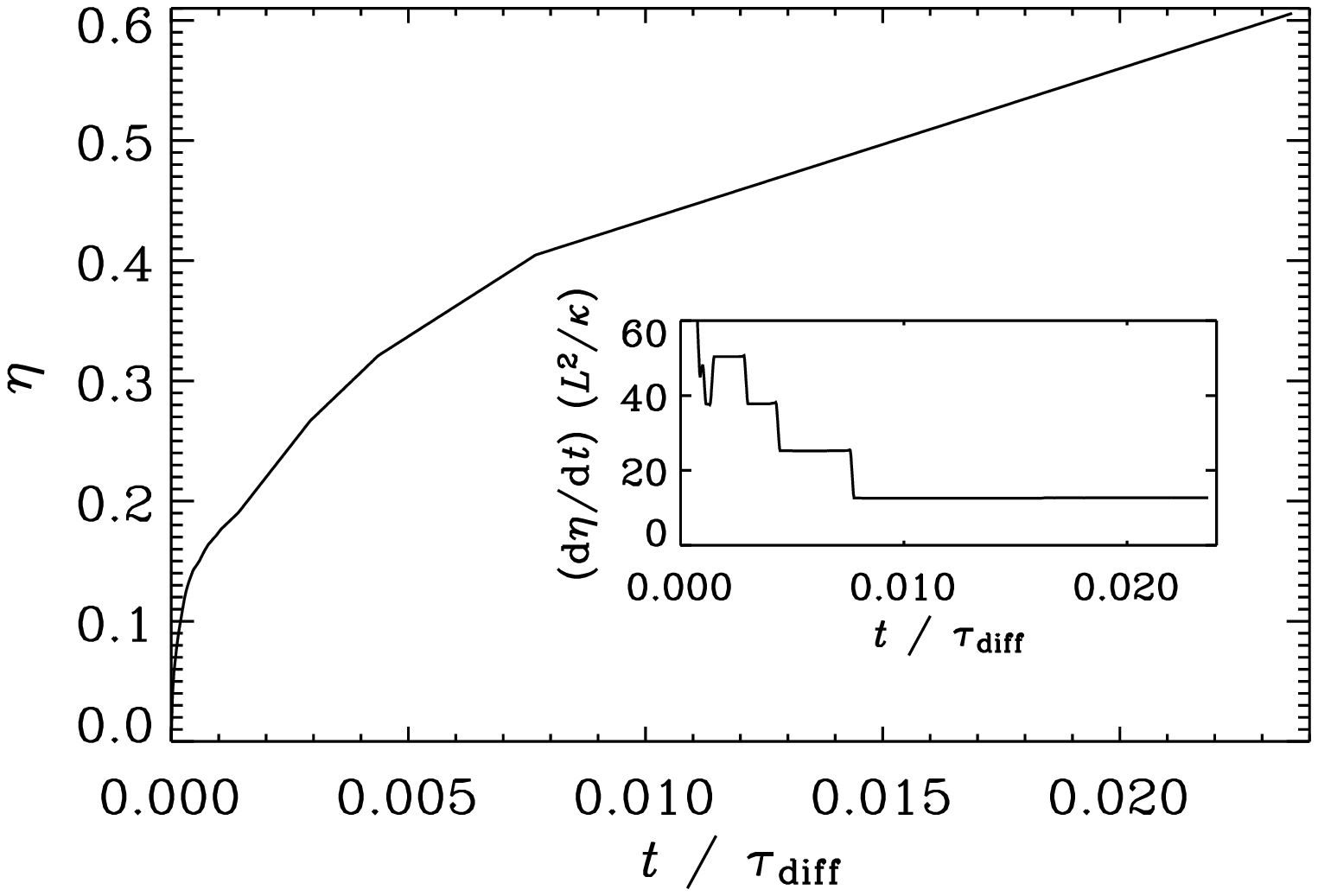}
\caption{
{\em Left}: Fractional concentration of one chirality versus position
($x$ and $y$) at four different times $t$, normalized by the diffusivity
$\kappa$ per total surface area $\lambda^2$, so $t\kappa/\lambda^2$
is nondimensional.
In this numerical simulation, $\kappa/(\lambda^2\lambda_0)=2\times10^{-4}$,
and the resolution was $1024^2$ mesh points.
The number of disconnected regions decreases from 4 in the
first plot to 3, 2, and 1.
{\em Right}: Evolution of enantiomeric excess $\eta$ for the model
shown in the left.
The inset shows the normalized slope.
Note the four distinct regimes with progressively decreasing slope.
Adapted from \cite{BM04}.
}\label{peta_pxxx_chiral}\end{figure}

\begin{figure}[t!]\begin{center}
\includegraphics[width=.45\columnwidth]{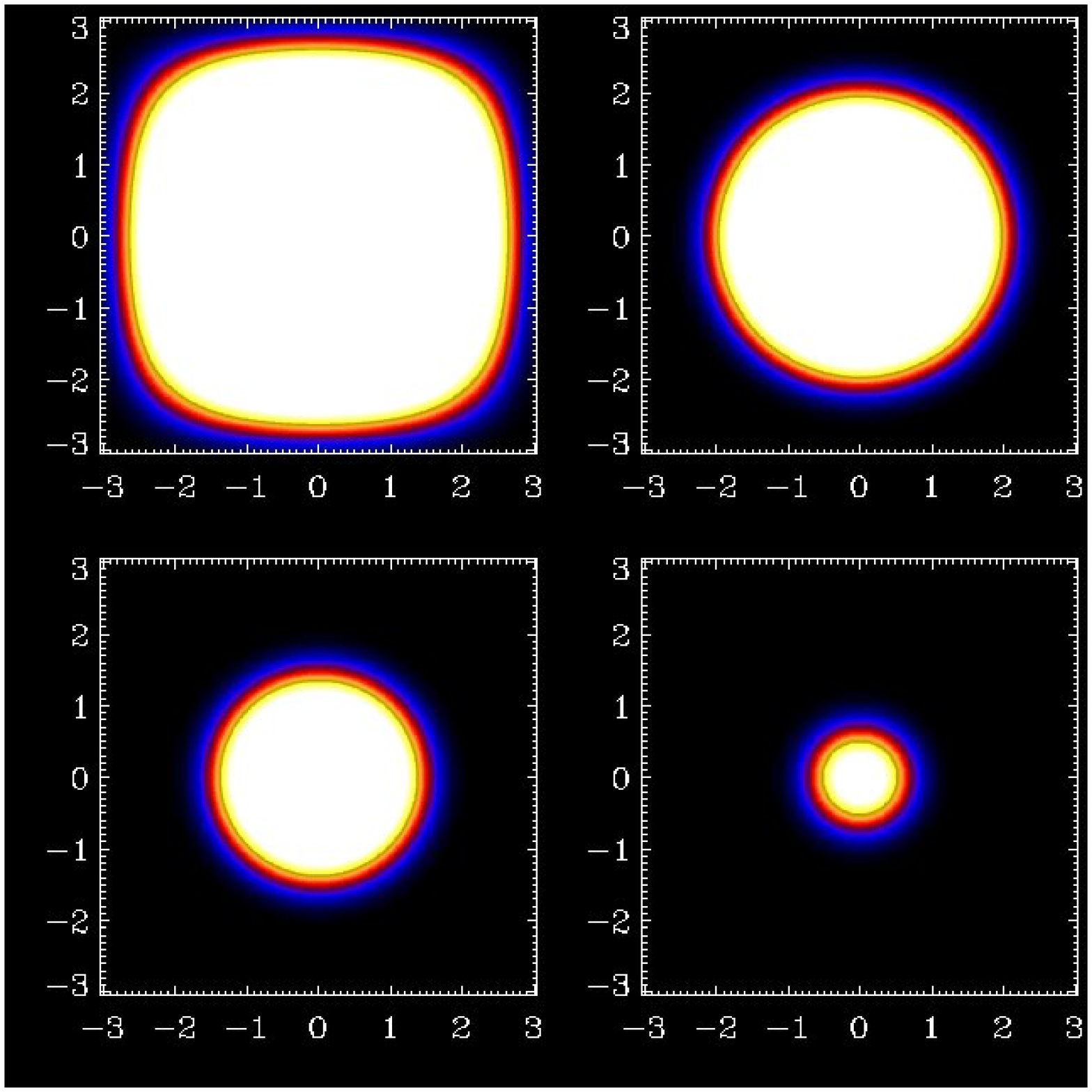}
\includegraphics[width=.54\columnwidth]{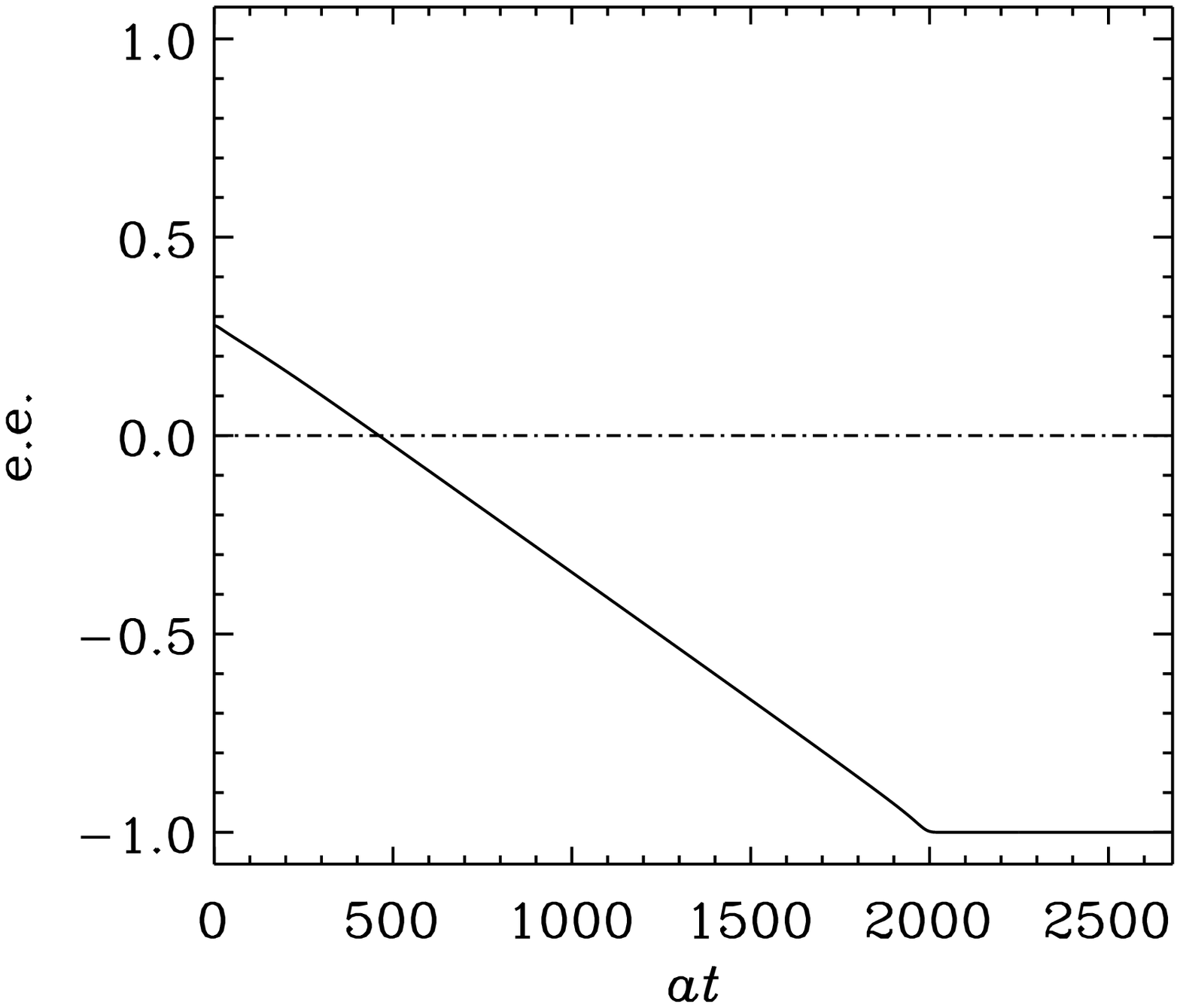}
\end{center}\caption[]{
{\em Left}: Shrinking of an initially large patch of molecules of the
\D form surrounded by molecules of the \L form.
{\em Right}: The resulting enantiomeric excess (e.e.) versus time $t$,
scaled with the activation rate $a$, so $at$ is nondimensional.
Note that it was initially positive, but reaches later complete
homochirality with $\eta=-1$.
}\label{Cpee}\end{figure}

More interesting dynamics is possible when the system is two-dimensional,
corresponding to different locations on the Earth's surface.
In that case, the fronts between regions with monomers and polymers of
opposite handedness can be curved.
It turned out that then the fronts are never precisely straight and can
therefore still propagate.
Interestingly, the propagation is always in the direction of maximum
curvature.
This result has also been obtained by \cite{GW12}.
This then implies that a closed circular front will always shrink and
never expand; see left \Fig{peta_pxxx_chiral}.
The speed of shrinking depends just on the number of individual closed fronts.
Each time a closed front merges with another one to form a single one,
the speed decreases; see the right panel of \Fig{peta_pxxx_chiral}.
In particular, this means that, even if, say, the \L enantiomers were
initially in the majority, but in such a way that they would be enclosed
in an island surrounded by enantiomers of the opposite handedness,
the enantiomeric excess would develop toward the handedness that was
present on the periphery of the domain;
see \Fig{Cpee} for such an illustration with the model of \cite{Plasson}.

% ln -s ~/tex/fabio/spatcomp/fig/C.ps C.eps
% ln -s ~/tex/fabio/spatcomp/fig/pee.ps -i pee.eps

\begin{figure}[t!]\begin{center}
\includegraphics[width=.4\textwidth]{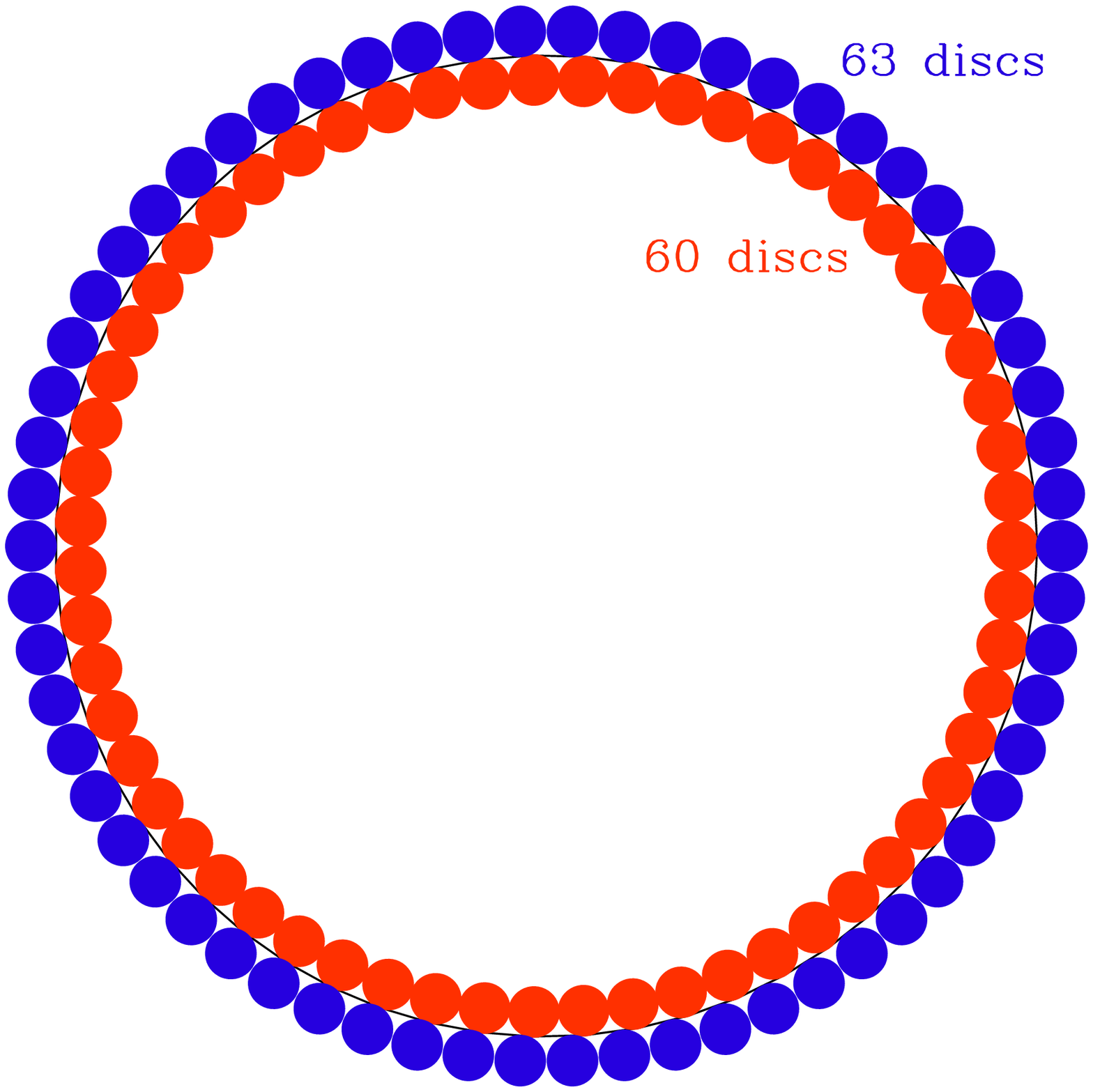}
\includegraphics[width=.4\textwidth]{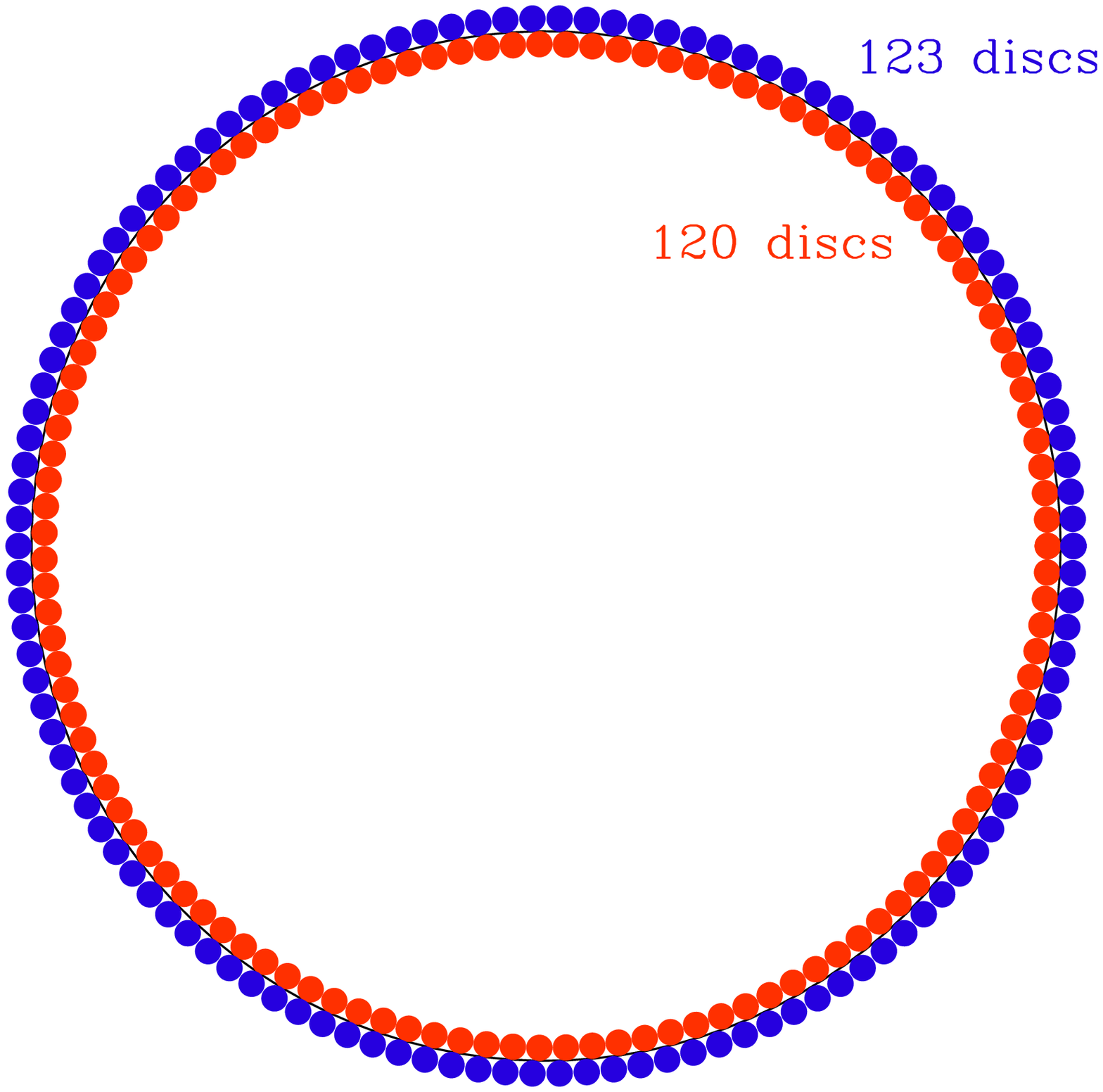}
\end{center}\caption[]{
Sketch illustrating that densely packed discs inside the periphery of
a circle differ in their number from those outside the periphery by just 3.
This result is independent of the total number; compare the left and
right illustrations with 60 and 120 discs, respectively.
As time goes on, pairs of red and blue discs get eliminated and
the circle shrinks, because the number of discs inside the periphery
is slightly smaller (by three) than the number of discs outside the
periphery.
The smallness of the difference in the numbers on the inner and outer
peripheries is the reason for the shrinking of the circle slowing down.
}\label{psketch}\end{figure}

The reason for this particular propagation direction is quite simple.
Imagine that we place \D and \L molecules around a circular front,
then the number of molecules on the inner front is would be one less
than the number of molecules on the outer front; see \Fig{psketch}
for an illustration.\footnote{
To understand why the difference in the number of molecules between the
outer and inner circles is always just three, let us imagine the molecules
being represented by little discs of radius $r$ on the periphery of a
circle of radius $R$.
The circumferences of the outer and inner peripheries are
$2\pi(R\pm r)=2\pi R\pm\pi r$ for the upper and lower signs, respectively.
The difference is therefore $2\pi r\approx3 d$, where $d=2r$ is the
diameter of each disc.
The difference in the number of discs is there for three.
}

\section{Recycling Frank: the peptide model of Plasson et al.}
\label{RecyclingFrank}

We said already in \Sec{EvidenceAutocatalysis} that autocatalysis may
not be a particularly evident process on the early Earth.
For that reason, \cite{Plasson} devised a completely different
mechanism that they advertised as ``recycling Frank''.
It is based on the combination of the following four important
reactions: activation (A), polymerization (P), epimerization (E), and
depolymerization (D).
So the resulting model is also referred to as the APED model.
A variant of this model was studied by \cite{KK18}.

It is important to emphasize that there is no explicit autocatalytic reaction.
However, the combined sequence of reactions \citep{BLL07}
\EQ
D+L
\stackrel{a}{\longrightarrow}D^*+L
\stackrel{p}{\longrightarrow}DL
\stackrel{e}{\longrightarrow}LL
\stackrel{h}{\longrightarrow}L+L,
\EN
\EQ
L+D
\stackrel{a}{\longrightarrow}L^*+D
\stackrel{p}{\longrightarrow}LD
\stackrel{e}{\longrightarrow}DD
\stackrel{h}{\longrightarrow}D+D.
\EN
does effectively result in an autocatalytic reaction, but it is not a
direct one.
First of all, it requires an activation step, indicated by asterisk,
a polymerization step (with the rate constant $p$),
an epimerization step (with the rate constant $e$),
and finally a depolymerization step (with the rate constant $h$).
Because the autocatalysis in indirect, this sequence of steps can
therefore be regarded as a simple example of a network catalysis
\citep{Plasson15,Hoc17}.

\section{Fluctuations instead of autocatalysis or enantiomeric cross-inhibition}

\begin{figure}[t!]\begin{center}
\includegraphics[width=\textwidth]{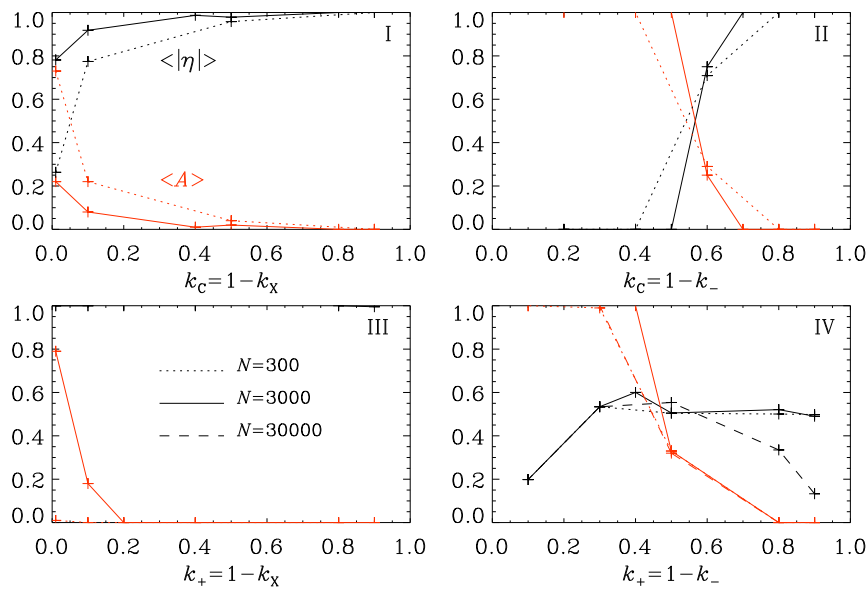}
\end{center}\caption[]{
Bifurcation diagrams of $\bra{|\eta|}$ (black) and $\bra{A}$ (red)
for $N=3000$ (solid lines) and $N=300$ (dotted lines)
as a function of parameters for models~I, II, III, and IV.
Adapted from \cite{Bra19}.
}\label{pmodels2}\end{figure}

During the last decade, there has been some increased interest in the
role of fluctuations; see a recent review by \cite{Wal17}.
Fluctuations can play important roles in diluted systems, in which
the number of molecules is small.
In such case, rate equations no longer provide a suitable description of
the relevant kinetics when the system is dilute and reactions are rare
\citep{Gil77,Tox14}.
In that case, a stochastic approach must be adopted.
This may be relevant to the work of \cite{Tox13}, where homochirality has
been found without apparent autocatalysis or enantiomeric cross-inhibition.
Instead of solving rate equations, as discussed in the previous sections,
one solves stochastic equations.
This means that at each reaction step, the state of the system changes,
but with a reaction that is taken to depend on chance with a certain
probability.
The system is then described by vector $\qq=(n_A,\,n_D,\,n_L)$, where
$n_A$ denotes the numbers of achiral molecules and $n_D$ and $n_L$
denotes the number of molecules of the \D and \L forms, respectively.
In the model of \cite{Bra19}, seven different reactions were considered,
each with a certain probability.
Not all those seven reactions need to be possible in a certain experiment,
so the probability for some reactions can be zero.
Applying a single reaction step with enantiomeric cross inhibition implies
\EQ
(n_A,\,n_D,\,n_L) \stackrel{k_\times~}{\longrightarrow} (n_A+2,\,n_D-1,\,n_L-1),
\EN
i.e., the numbers of $D$ and $L$ get reduced by one, and that of $A$
increases by two.
We can also include spontaneous deracemization reactions, i.e.,
\EQA
(n_A,\,n_D,\,n_L) \stackrel{k_+~}{\longrightarrow} (n_A-1,\,n_D+1,\,n_L),\\
\label{kplus1}
(n_A,\,n_D,\,n_L) \stackrel{k_+~}{\longrightarrow} (n_A-1,\,n_D,\,n_L+1).
\label{kplus2}
\ENA
To model different reaction rates, the different reactions must happen
with different probabilities.
This is done by taking at each reaction step a random number between
zero and one.
Suppose we want to model enantiomeric cross inhibition together with
spontaneous deracemization, then the probability that the first reaction
happens is proportional to $k_\times$, and the probability that one of
the other two reactions in \Eqs{kplus1}{kplus2} occurs is proportional
to $k_+/2$.
If we also allow for the possibility that nothing happens (probability
proportional to $k_0$), then our scheme with $\qq\to\qq+\Delta\qq$ is as follows:
\begin{align}
&\mbox{if} \quad 0  \leq r < r_1\equiv k_\times  && \mbox{then $\Delta\qq=(2,-1,-1)$},\\
&\mbox{if} \quad r_1\leq r < r_2\equiv r_1+k_+/2 && \mbox{then $\Delta\qq=(-1,1,0)$},\\
&\mbox{if} \quad r_2\leq r < r_3\equiv r_2+k_+/2 && \mbox{then $\Delta\qq=(-1,0,1)$},\\
&\mbox{if} \quad r_3\leq r < 1                   && \mbox{then $\Delta\qq=\bm{0}$ (no reaction)}.
\end{align}
Note that $k_\times+k_++k_0=1$ is here assumed.
This particular experiment was referred to as experiment III in
\cite{Bra19}, where $k_0=0$ was assumed.
As he varied $k_\times$, $k_+$ was assumed to vary
correspondingly such that $k_+=1-k_\times$.
The results of this experiment are similar to those with spontaneous
deracemization replaced by autocatalysis, which is referred
to as experiment~I in \Fig{pmodels2}.
Here, the autocatalysis rate is varied such that $k_{\rm C}=1-k_\times$.
This is the standard Frank model, but for a diluted system, while
model~III is close to that of \cite{SHS08,SHS09}, who were the first
to find a transition to full homochirality even without autocatalysis.
Next, in experiment~II, there is autocatalysis, but no enantiomeric
cross inhibition and just spontaneous racemization instead.
This type of model was first considered by \cite{JBG15,JBG17}.
The transition to full homochirality was originally though impossible
in such a model \citep{Stich}.

In a comparative study, all these processes were studied within a single
unified model.
In \Fig{pmodels2} we show the results of the four different experiments.
We mentioned already experiments~I--III.
In experiment~IV, by comparison, there is just racemization and deracemization,
but neither autocatalysis nor enantiomeric cross inhibition.
In that case, the average of the modulus of the enantiomeric excess,
$\bra{|\eta|}$, no longer reaches unity, but levels off at about $0.5$
when $k_+\ga0.4$.
We also see from the red lines that the achiral compounds ($A$) get
depleted in favor of producing chiral ones either of the \D or the \L form.

\section{Chirality from a Martian Labeled Release experiment}

Back in 1976, when the Viking~I and II landers visited the Chryse Planitia
and Utopia Planitia regions, respectively, many of the things we now know
about Mars were still unclear.
In particular, the existence of water on Mars was still very much an
open question.
Nevertheless, one was relatively optimistic at the time.
Both landers came with advanced experiments on board to look for life.
One of the experiments, the Labeled Release (LR) experiment, was actually
successful \citep{LS76,LS77}, but another experiment never detected any
organics, which was decisive enough to conclude that no life was detected
after all \citep{Klein76}.

The idea behind the LR experiment is simple: take Martian soil, mix it
with water and organics as nutrients, and see whether a metabolic
reaction occurs that decomposes the nutrients and produces a gaseous
waste product, for example methane or carbon dioxide; see the recent
account by \cite{LS16}, where detailed tests with various terrestrial
soils were presented.
The carbon atoms of the nutrients were labelled with carbon-14 isotopes,
a technique commonly used in medicine, which allows one to trace those
labeled carbon atoms by their radioactivity.
To identify the gaseous waste product, one simply measured the level
of radioactivity.
Control experiments with sterilized soil showed that only fresh
Martian soil produced a reaction.
The Viking laboratories were flexible enough to perform additional
experiments with lower sterilization temperatures.
The critical temperature below which no sterilization occurred
was found to be around $50^\circ$.
Those temperatures would appear reasonable for Martian cryophiles,
but are generally too low for sterilization on Earth.
The experiment was tested in various deserts on Earth and it was able
to detect metabolism at measurable levels.

It is only since 2012 that organics were detected on the Martian surface
by the Curiosity rover; see \cite{Voosen} for a popular account.
We also know that organics get quickly destroyed by perchlorates, in
particular KClO$_4$, which were discovered on the Martian surface by
the Phoenix lander in 2008 \citep{Hec09}.
Such processes could potentially result in reactions found
with the LR experiment \citep{Quinn}, but it remains puzzling why a
critical sterilization temperature of $50^\circ$ was found, and
not much higher, for example.
Thus, while an explanation in terms of abiotic processes has not been
fully conclusive \citep{Valdivia-Silva}, the explanation that life was
actually detected might seem more straightforward \citep{LS16}.
However, as already noted by Carl Sagan, ``the more extraordinary
the claim, the more extraordinarily well-tested the evidence must be.
The person making the extraordinary claim has the burden of proving to
the experts at large that his or her belief has more validity than the
one almost everyone else accepts.''
In any case, it seems justified to repeat this experiment to clarify
the phenomenon that the Viking landers discovered back in 1976; see
also the recent paper by \cite{Carrier}.

\begin{figure}[t!]\begin{center}
\includegraphics[width=\textwidth]{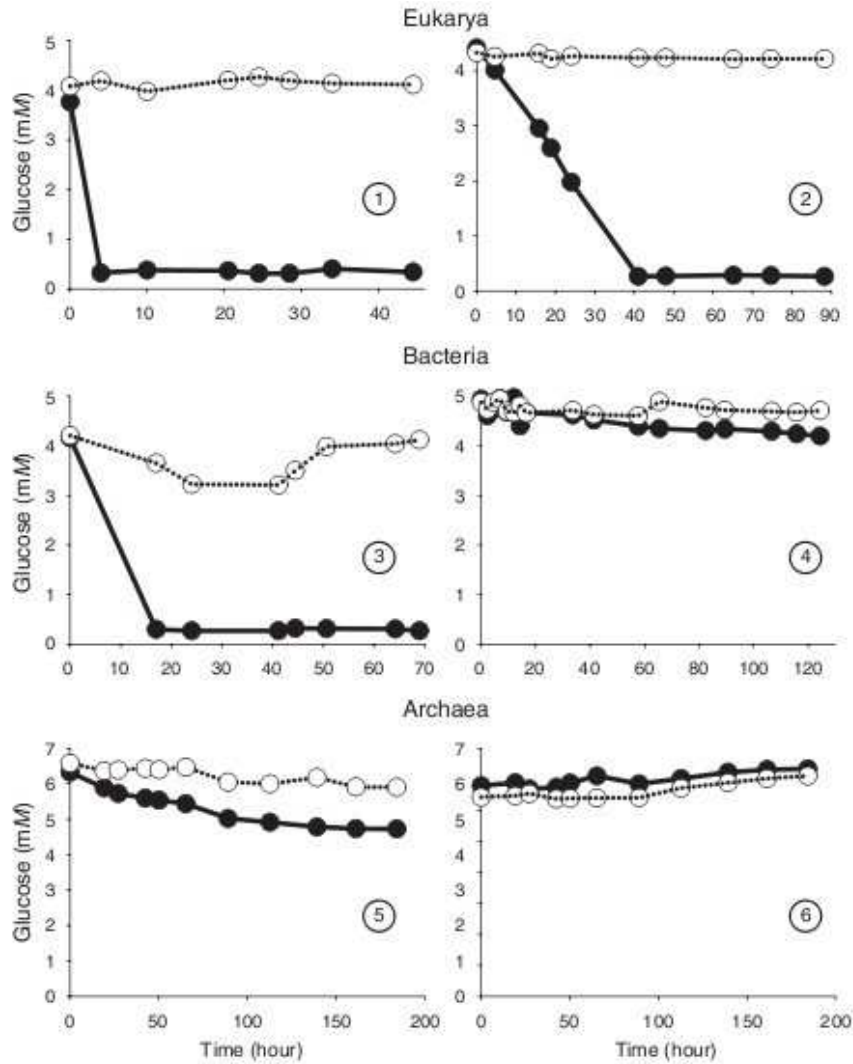}
\end{center}\caption[]{
Metabolic consumption of $\D$-glucose (filled symbols) and $\L$-glucose
(open symbols) and by (1) {\em Saccharomyces cerevisiae}, (2) {\em
Penicilium expansum}, (3) {\em E. coli}, (4) {\em Micrococcus luteus},
(5) {\em Natronobacterium sp.}, and (6) {\em Halostagnicola sp.}
Adapted from \cite{Sun09}.
}\label{FSun09}\end{figure}

Given that the experiment is relatively simple and can detect life under
harsh conditions on Earth, it would be interesting to repeat some
variants of it in the future.
One such variant would be to allow for the detection of handedness.
This would constitute a more conclusive signature of life than just the
discovery of some metabolism.
Such an experiment can be done by using chiral nutrients, which goes back
to the old findings of Pasteur of 1857 that certain microorganisms had
a preference for consuming $(+)$-tartaric acid over $(-)$-tartaric acid;
see the reviews by \cite{Gal08} and \cite{Sevin}.

In \Fig{FSun09} we show the result of recent experiments by \cite{Sun09}
using different types of eukarya, bacteria, and archaea, which were
given either \D sugars or \L sugars.
In most of the cases there was a clear preference in the microbes taking
up the naturally occurring \D sugars compared with the synthetically
produced \L sugars.
Subsequent work showed that the specificity for some microbes is low
and that some of those can use sugars of the opposite chirality also
\citep{Moazeni}.
Although the dependence on the type of nutrients has not yet been studied
in detail, it may be important to allow for a broad range of different ones
in an attempt to account for such ambiguities.

\section{Conclusions}

Louis Pasteur was well ahead of his time when he identified the
biological role of chirality in living matter.
Particular remarkable is his realization that, during fermentation,
the metabolic uptake of nutrients of opposite chiralities is different.
To understand why this experiment was not put in the context of
extraterrestrial life detection, we have to realize that in those years,
it was not uncommon to think of extraterrestrial life on Mars.
When the astronomer \cite{Herschel} discovered seasons on Mars, he wrote
in his paper in the Philosophical Transactions of the Royal Society
that this ``planet has a considerable but moderate atmosphere so that its
inhabitants probably enjoy a situation in many respects similar to ours.''
So, not just the existence of life, but the existence of {\em intelligent}
of life on Mars was commonly expected.
This only changed in 1964, when Mariner~4 returned the first flyby
pictures of Mars, which suggested that any life there would probably
only be of microbial nature.
But that Vikings~1 and 2 would not even find any organics on Mars was such
a shock to many that the search for life in the Universe appeared fairly
hopeless, and Mars exploration was put on hold for the next two decades.
This all changed since the turn of the century with the discovery of
extremophiles on Earth and the realization that terrestrial life has
existed since the time that stable continents existed.
Gradually, with the conclusive detection of water on Mars, the search
for extinct or extant life on Mars restarted, and Pasteur's discovery
of different metabolic uptakes of \D and \L nutrients may finally turn
into an actual Martian experiment.
As shown by \cite{Sun09}, this property can be used to detect the presence
of homochirality through in situ experiments.
Homochirality can also be detected through remote sensing by looking
for circular polarization.
This approach has been persued by \cite{LtenKate}, who found negative
or left-handed circularly polarized light emitted from terrestrial plant
life at about $680\,$nm; see also \cite{Avn21} for a recent review.
To what extent this technique can be used as a biomarker still needs
to be seen, but it is amazing to see once more how Pasteur's early
discoveries have shaped some important aspects of astrobiology.

While homochirality remains a property that is strongly associated with
life---or at least some chemical process that keeps the system far from
equilibrium---it is not clear whether we should expect it to have the
same or the opposite handedness as an Earth \citep{Bada96}.
To answer this question, one would need to have more realistic models
with meaningful estimates for the concentrations of suitable chemicals
in some protocells.
This would allow for an estimate of the level of fluctuations in relation
to the strengths of the small but systematic effects resulting from the
weak force.
It would be the only way of guaranteeing that each genesis of life
always produces the same chirality.
But for now, we should be satisfied if one could find (and understand)
any type of an extraterrestrial metabolic process that works differently
for nutrients of \D and \L forms.
Thus, the discussed possibilities would need to be put
on a quantitatively meaningfully basis.
And if it is not life, it certainly is interesting enough to deserve
serious attention!

\section*{Acknowledgements}

I am grateful to the two referees for many useful
comments and suggestions that have led to improvements of the manuscript.
In particular, I wish to acknowledge Rapha\"el's clarifications on the
differences in the terminology in terms of levorotatory and dextrorotatory
on the one hand and that in terms of \D and \L on the other.
This work was supported in part through the Swedish Research Council,
grant 2019-04234.
I acknowledge the allocation of computing resources provided by the
Swedish National Allocations Committee at the Center for Parallel
Computers at the Royal Institute of Technology in Stockholm.

%r e f


\begin{thebibliography}{}

\bibitem[Anders et al.(1964)]{Orgueil64}
Anders, E., DuFresne, E. R., Hayatsu, R., Cavaille, A., DuFresne, A., \& Fitch, F. W.\ysci{1964}{146}{1157}
{1161}{Contaminated meteorite}

\bibitem[Athavale et al.(2020)]{Athavale}
Athavale, S. V., Simon, A., Houk, K. N., \& Denmark, S. E.\yjour{2020}{Nat. Chem.}{12}{412}
{423}{Demystifying the asymmetry-amplifying, autocatalytic behaviour of the Soai reaction through structural, mechanistic and computational studies}

\bibitem[Avetisov et al.(1991)]{AGK91}
Avetisov, V. A., Goldanskii, V. I., \& Kuz'min, V. V.\yjour{1991}{Phys. Today}{44}{33}
{41}{Handedness, origin of life and evolution}

\bibitem[Avnir(2021)]{Avn21}
Avnir, D.\yjourN{2021}{New Astron. Rev.}{92}{101596}
{Critical review of chirality indicators of extraterrestrial life}

\bibitem[Baaske et al.(2007)]{Braun}
Baaske, P., Weinert, F. M., Duhr, S., Lemke, K. H., Russell, M. J., \& Braun D.\ypnas{2007}{104}{9346}
{9351}{Extreme accumulation of nucleotides in simulated hydrothermal pore systems}

\bibitem[Bada(1995)]{Bada95}
Bada, J. L.\ynat{1995}{374}{594}
{595}{Origins of homochirality}

\bibitem[Bada(1996)]{Bada96}
Bada, J. L.\yoleb{1996}{26}{518}
{519}{Amino acid homochirality on Earth and Mars}

\bibitem[Bailey(2001)]{Bailey01}
Bailey, J.\yoleb{2001}{31}{167}
{183}{Astronomical sources of circularly polarized light and the origin
of homochirality}

\bibitem[Bailey et al.(1998)]{Bailey98}
Bailey, J., Chrysostomou, A., Hough, J. H., Gledhill, T. M., McCall, A.,
Clark, S., M\'enard, F., \& Tamura, M.\ysci{1998}{281}{672}
{674}{Circular polarization in star forming regions: implications for
biomolecular homochirality}

\bibitem[Barge et al.(2017)]{Bar17}
Barge, L. M., Branscomb, E., Brucato, J. R., Cardoso, S. S. S., Cartwright, J. H. E., Danielache, S. O., Galante, D., Kee, T. P., Miguel, Y., Mojzsis, S., Robinson, K. J., Russell, M. J., Simoncini, E., Sobron, P.\yoleb{2017}{47}{39}
{56}{Thermodynamics, disequilibrium, evolution: Far-from-equilibrium geological and chemical considerations for origin-of-life research}

\bibitem[Barge et al.(2019)]{Bar19}
Barge, L. M., Flores, E., Baum, M. M., VanderVelde, D. G., Russell, M. J.\ypnas{2019}{116}{4828}
{4833}{Redox and pH gradients drive amino acid synthesis in iron oxyhydroxide mineral systems}

\bibitem[Bonner et al.(1981)]{Bon81}
Bonner, W. A., Blair, N. E., \& Dirbas, F. M.\yol{1981}{11}{119}
{134}{Experiments on the abiotic amplification of optical activity}

\bibitem[Bonner(1991)]{Bon91}
Bonner, W. A.\yoleb{1991}{21}{59}
{111}{The origin and amplification of biomolecular chirality}

\bibitem[Bonner(2000)]{Bon00}
Bonner, W. A.\yjour{2000}{Chirality}{12}{114}
{126}{Parity violation and the evolution of biomolecular homochirality}

\bibitem[Boyd et al.(2018)]{Boyd}
Boyd, R. N., Famiano, M. A., Onaka, T., \& Kajino, T.\yapjN{2018}{856}{26}
{Sites that can produce left-handed amino acids in the supernova neutrino amino acid processing model}

\bibitem[Brandenburg(2019)]{Bra19}
Brandenburg, A.\yoleb{2019}{49}{49}
{60}{The limited roles of autocatalysis and enantiomeric cross-inhibition in achieving homochirality in dilute systems}

\bibitem[Brandenburg(2020)]{Bra20cor}
Brandenburg, A.\yjour{2020}{Infectious Disease Modelling}{5}{681}
{690}{Piecewise quadratic growth during the 2019 novel coronavirus epidemic}

\bibitem[Brandenburg et al.(2005)]{BAHN}
Brandenburg, A., Andersen, A. C., H\"ofner, S., \& Nilsson, M.\yoleb{2005}{35}{225}
{241}{Homochiral growth through enantiomeric cross-inhibition}

\bibitem[Brandenburg et al.(2007)]{BLL07}
Brandenburg, A., Lehto, H. J., \& Lehto, K. M.\yab{2007}{7}{725}
{732}{Homochirality in an early peptide world}

\bibitem[Brandenburg \& Multam\"aki(2004)]{BM04}
Brandenburg, A., \& Multam\"aki, T.\yijaS{2004}{3}{209}
{219}{How long can left and right handed life forms coexist?}

\bibitem[Braun \& Libchaber(2002)]{Braun02}
Braun, D., \& Libchaber, A.\yprlN{2002}{89}{188103}
{Trapping of DNA by thermophoretic depletion and convection}

\bibitem[Carrier et al.(2020)]{Carrier}
Carrier, B. L., Beaty, D. W., Meyer, M. A., Blank, J. G., Chou, L., DasSarma, S., Des Marais, D. J., Eigenbrode, J. L., Grefenstette, N., Lanza, N. L., Schuerger, A. C., Schwendner, P., Smith, H. D., Stoker, C. R., Tarnas, J. D., Webster, K. D., Bakermans, C., Baxter, B. K., Bell, M. S., Benner, S. A., et al.\yab{2020}{20}{785}
{814}{Mars extant life: What's next? Conference report}

\bibitem[Cech(1986)]{Cech}
Cech, T. R.\ypnas{1986}{83}{4360}
{4363}{A model for the RNA-catalyzed replication of RNA}

\bibitem[Cloez(1864)]{Cloez}
Cloez, S.\yjourN{1864}{Compt. Rend. Acad. Sci. Paris}{58}{986}
{Note sur la composition chimique de la pierre}

\bibitem[Davies \& Lineweaver(2005)]{DL05}
Davies, P. C. W. \& Lineweaver, C. H.\yab{2005}{5}{154}
{163}{Finding a second sample of life on Earth}

\bibitem[Derewenda(2008)]{Der08}
Derewenda, Z. S.\yjour{2008}{Acta Cryst.}{A64}{246}
{258}{On wine, chirality and crystallography}

\bibitem[Ehrenfreund et al.(2001)]{Ehrenfreund}
Ehrenfreund, P., Glavin, D. P., Botta, O., Cooper, G., \& Bada, J. L.\ypnas{2001}{98}{2138}
{2141}{Extraterrestrial amino acids in Orgueil and Ivuna: Tracing the parent body of CI type carbonaceous chondrites}

\bibitem[Engel \& Macko(1997)]{EM97}
Engel, M. H., \& Macko, S. A.\ynat{1997}{389}{265}
{268}{Isotopic evidence for extraterrestrial non-racemic amino acids in
the Murchison meteorite}

\bibitem[Fajszi \& Cz\'eg\'e(1981)]{FC81}
Fajszi, Cs. \& Cz\'eg\'e, J.\yol{1981}{11}{143}
{162}{Critical evaluation of mathematical models for the amplification of chirality}

\bibitem[Frank(1953)]{Frank}
Frank, F. C.\yjour{1953}{Biochim.\ Biophys.\ Acta}{11}{459}
{464}{On spontaneous asymmetric synthesis}

\bibitem[Gal(2008)]{Gal08}
Gal, J.\yjour{2008}{Chirality}{20}{5}
{19}{The discovery of biological enantioselectivity: Louis Pasteur and the fermentation of tartaric acid, 1857--A review and analysis 150 yr later}

\bibitem[Gehring et al.(2010)]{Gehring}
Gehring, T., Busch, M., Schlageter, M., \& Weingand, D.\yjour{2010}{Chirality}{22}{E173}
{E182}{A concise summary of experimental facts about the Soai reaction}

\bibitem[Gilbert(1986)]{Gil86}
Gilbert, W.\ynat{1986}{319}{618}
{618}{Origin of life -- the RNA world}

\bibitem[Gillespie(1977)]{Gil77}
Gillespie, D. T.\yjour{1977}{J. Phys. Chem.}{81}{2340}
{2361}{Exact stochastic simulation of coupled chemical reactions}

\bibitem[Gleiser \& Walker(2012)]{GW12}
Gleiser, M., \& Walker, S. I.\yija{2012}{11}{287}
{296}{Life's chirality from prebiotic environments}

\bibitem[Globus \& Blandford(2020)]{GB20}
Globus, N., \& Blandford, R. D.\yapjlN{2020}{895}{L11}
{The chiral puzzle of life}

\bibitem[Goldanskii \& Kuz'min(1989)]{GK89}
Goldanskii, V. I., \& Kuz'min, V. V.\yjour{1989}{Sov. Phys. Usp.}{32}{1}
{29}{Spontaneous breaking of mirror symmetry in nature and the origin of life}

\bibitem[Goldhaber et al.(1957)]{Goldhaber}
Goldhaber, M., Grodzins, L., \& Sunyar, A. W.\ypr{1957}{106}{826}
{828}{Evidence for circular polarization of bremsstrahlung produced by beta rays}

\bibitem[Guerrier-Takada \& Altman(1984)]{Altman}
Guerrier-Takada, C., \& Altman, S.\ysci{1984}{223}{285}
{286}{Catalytic activity of an RNA molecule prepared by transcription in vitro}

\bibitem[Hecht et al.(2009)]{Hec09}
Hecht, M. H., Kounaves, S. P., Quinn, R. C., West, S. J., Young, S. M. M., Ming, D. W., Catling, D. C., Clark, B. C., Boynton, W. V., Hoffman, J., DeFlores, L. P., Gospodinova, K., Kapit, J., Smith, P. H.\ysci{2009}{325}{64}
{67}{Detection of Perchlorate and the Soluble Chemistry of Martian Soil at the Phoenix Lander Site}

\bibitem[Hegstrom(1984)]{Hegstrom}
Hegstrom, R. A.\yjour{1984}{Orig. Life}{14}{405}
{414}{Parity nonconservation and the origin of biological chirality -- theoretical calculations}

\bibitem[Hegstrom et al.(1980)]{HRS80}
Hegstrom, R. A., Rein, D. W., \& Sandars, P. G. H.\yjour{1980}{J. Chem. Phys.}{73}{2329}
{2341}{Calculation of the parity nonconserving energy difference between mirror-image molecules}

\bibitem[Herschel(1784)]{Herschel}
Herschel, W.\yptrs{1784}{74}{233}
{273}{On the remarkable appearances at the polar regions of the planet Mars, the inclination of its axis, the position of its poles, and its spheroidical figure; with a few hints relating to its real diameter and atmosphere}

\bibitem[Hochberg et al.(2017)]{Hoc17}
Hochberg, D., Bourdon G., Rub\'en D., \'Agreda B, Jes\'us A., \& Rib\'o, J. M.\yjour{2017}{Phys. Chem. Chem. Phys.}{19}{17618}
{17636}{Stoichiometric network analysis of spontaneous mirror symmetry breaking in chemical reactions}

\bibitem[Jafarpour et al.(2015)]{JBG15}
Jafarpour, F., Biancalani, T., Goldenfeld, N.\yprlN{2015}{115}{158101}
{Noise-induced mechanism for biological homochirality of early life self-replicators}

\bibitem[Jafarpour et al.(2017)]{JBG17}
Jafarpour, F., Biancalani, T., \& Goldenfeld, N.\ypreN{2017}{95}{032407}
{Noise-induced symmetry breaking far from equilibrium and the emergence of biological homochirality}

\bibitem[Joyce et al.(1984)]{Joyce84}
Joyce, G. F., Visser, G. M., van Boeckel, C. A. A., van Boom, J. H.,
Orgel, L. E., \& Westrenen, J.\ynat{1984}{310}{602}
{603}{Chiral selection in poly(C)-directed synthesis of oligo(G)}

\bibitem[K\"all\'en et al.(1985)]{KAM85}
K\"all\'en, A., Arcuri, P., \& Murray, J. D.\yjour{1985}{J. Theor. Biol.}{116}{377}
{393}{A simple model for the spatial spread and control of rabies}

\bibitem[Klein et al.(1976)]{Klein76}
Klein, H. P., Horowitz, N. H., Levin, G. V., Oyama, V. I., Lederberg, J., Rich, A., Hubbard, J. S., Hobby, G. L., Straat, P. A., Berdahl, B. J., Carle, G. C., Brown, F. S., \& Johnson, R. D.\ysci{1976}{194}{99}
{105}{The Viking biological investigation: Preliminary results}

\bibitem[Kondepudi \& Nelson(1983)]{KN83}
Kondepudi, D. K., and Nelson, G. W.\yprl{1983}{50}{1023}
{1026}{Chiral symmetry breaking in nonequilibrium systems}

\bibitem[Kondepudi \& Nelson(1985)]{KN85}
Kondepudi, D. K., \& Nelson, G. W.\ynat{1985}{314}{438}
{441}{Weak neutral currents and the origin of biomolecular chirality}

\bibitem[Konstantinov \& Konstantinova(2018)]{KK18}
Konstantinov, K. K., \& Konstantinova, A. F.\yoleb{2018}{48}{93}
{122}{Chiral symmetry breaking in peptide systems during formation of life on Earth}

\bibitem[Lee \& Yang(1956)]{LY56}
Lee, T. D., \& Yang, C. N.\ypr{1956}{104}{254}
{258}{Question of parity conservation in weak interactions}

\bibitem[Levin(2009)]{Lev09}
Levin, G. V.\yabN{2009}{9}{503}
{Comment on ``Stereo-specific glucose consumption may be used to distinguish between chemical and biological reactivity on Mars: A preliminary test on Earth''}

\bibitem[Levin \& Straat(1976)]{LS76}
Levin, G. V., \& Straat, P. A.\yoleb{1976}{7}{293}
{311}{Labeled Release -- Experiment in radiorespirometry}

\bibitem[Levin \& Straat(1977)]{LS77}
Levin, G. V., \& Straat, P. A.\yjgr{1977}{82}{4663}
{4667}{Recent results from the Viking Labeled Release Experiment on Mars}

\bibitem[Levin \& Straat(2016)]{LS16}
Levin, G. V., \& Straat, P. A.\yab{2016}{16}{798}
{810}{The case for extant life on Mars and its possible detection by the Viking Labeled Release Experiment}

\bibitem[Mason \& Tranter(1984)]{MT84}
Mason, S. F., \& Tranter, G. E.\yjour{1984}{Mol. Phys.}{53}{1091}
{1111}{The parity-violating energy difference between enantiomeric molecules}

\bibitem[McVoy(1957)]{McVoy}
McVoy, K. W.\ypr{1957}{106}{828}
{829}{Circular polarization of bremsstrahlung from polarized electrons in Born approximation}

\bibitem[Meierhenrich \& Thiemann(2004)]{MT04}
Meierhenrich, U. J., \& Thiemann, W. H.-P.\yoleb{2004}{34}{111}
{121}{Photochemical concepts on the origin of biomolecular asymmetry}

\bibitem[Miller(1953)]{Mil53}
Miller, S. L.\ysci{1953}{117}{528}
{529}{A production of amino acids under possible primitive Earth conditions}

\bibitem[Moazeni et al.(2010)]{Moazeni}
Moazeni, F., Zhang, G., \& Sun, H. J.\yab{2010}{10}{397}
{402}{Imperfect asymmetry of life: Earth microbial communities prefer D-lactate but can use L-lactate also}

\bibitem[Murray et al.(1986)]{Mur86}
Murray, J. D., Stanley, E. A., \& Brown, D. L.\yjour{1986}{Proc. Roy. Soc. Lond. Ser. B}{229}{111}
{150}{On the spatial spread of rabies among foxes}

\bibitem[Noble(1974)]{Nob74}
Noble, J. V.\yjour{1974}{Nature}{250}{726}
{728}{Geographic and temporal development of plagues}

\bibitem[Pasteur(1853)]{Pasteur1853}
Pasteur, L.\yjour{1853}{Ann. Phys.}{166}{504}
{509}{Umwandlung der Weins\"aure in Traubens\"aure. Entdeckung von unwirksamer Weins\"aure. Neue Methode der Zerlegung von Traubens\"aure in Rechts- und in Linksweins\"aure}

\bibitem[Pasteur(1922)]{Masson}
Pasteur, L.\yproc{1922}{56}
{99}{Recherches sur les propri\'et\'es sp\'ecifiques des deux acides qui composent l'acide rac\'emique}
{Annales de chimie et physique}{Paris: Masson}{3rd edition, XXVIII}
%printed in #uvres de Pasteur, volume 1, Dissymétrie moléculaire, Masson (1922), online at Gallica (the figure is on p.90).

\bibitem[Patty et al.(2019)]{LtenKate}
Patty, C. H. L., ten Kate, I. L., Buma, W. J., van Spanning, R. J. M., Steinbach, G., Ariese, F., \& Snik, F.\yab{2019}{19}{1221}
{1229}{Circular spectropolarimetric sensing of vegetation in the field: possibilities for the remote detection of extraterrestrial life}

\bibitem[Pizzarello \& Cronin(2000)]{PC00}
Pizzarello, S., \& Cronin, J. R.\yjour{2000}{Geochimica et Cosmochimica Acta}{64}{329}
{338}{Non-racemic amino acids in the Murray and Murchison meteorites}

\bibitem[Plasson et al.(2004)]{Plasson}
Plasson, R., Bersini, H., \& Commeyras, A.\ypnas{2004}{101}{16733}
{16738}{Recycling Frank: spontaneous emergence of homochirality in
noncatalytic systems}

\bibitem[Plasson(2015)]{Plasson15}
Plasson, R.\yprocN{2015}{483}
{Chemical reaction network}
{Encyclopedia of Astrobiology}
{Gargaud, M., et al.}
{Springer Berlin Heidelberg}

\bibitem[Quinn et al.(2005)]{Quinn}
Quinn, R. C., Zent, A. P., Grunthaner, F. J., Ehrenfreund, P., Taylor, C. L., \& Garry, J. R. C.\yjour{2005}{Planet. Space Sci.}{53}{1376}
{1388}{Detection and characterization of oxidizing acids in the Atacama Desert using the Mars oxidation instrument}

\bibitem[Rothery et al.(2008)]{RGS08}
Rothery, D. A., Gilmour, I., \& Sephton, M. A.\ybook{2008}
{An Introduction to Astrobiology}{Cambridge University Press}

\bibitem[Russell(2006)]{Rus06}
Russell, M.\yjour{2006}{Am. Sci.}{94}{32}
{39}{First life}

\bibitem[Russell et al.(2014)]{Rus14}
Russell, M. J., Barge, L. M., Bhartia, R., Bocanegra, D., Bracher, P. J., Branscomb, E., Kidd, R., McGlynn, S., Meier, D. H., Nitschke, W., Shibuya, T., Vance, S., White, L., Kanik, I.\yab{2014}{14}{308}
{343}{The drive to life on wet and icy worlds}

\bibitem[Sandars(2003)]{San03}
Sandars, P. G. H.\yoleb{2003}{33}{575}
{587}{A toy model for the generation of homochirality during polymerization}

\bibitem[Sandars(2005)]{San05}
Sandars, P. G. H.\yija{2005}{4}{49}
{61}{Chirality in the RNA world and beyond}

\bibitem[Sevin(2015)]{Sevin}
Sevin, A.\yjour{2015}{Bibnum [Online], Chimie}{459}{1}
{10}{Pasteur and molecular chirality}, 
%http://journals.openedition.org/bibnum/459

\bibitem[Soai et al.(1995)]{Soai}
Soai, K., Shibata, T., Morioka, H., \& Choji, K.\ynat{1995}{378}{767}
{768}{Asymmetric autocatalysis and amplification of enantiomeric excess
of a chiral molecule}

\bibitem[Stich et al.(2016)]{Stich}
Stich, M., Rib\'o, Josep M., Blackmond, D. G., \& Hochberg, D.\yjourN{2016}{J. Chem. Phys.}{145}{074111}
{Necessary conditions for the emergence of homochirality via autocatalytic self-replication}

\bibitem[Sugimori et al.(2008)]{SHS08}
Sugimori, T., Hyuga, H., Saito, Y.\yjourN{2008}{J. Phys. Soc. Jap.}{77}{064606}
{Fluctuation induced homochirality}

\bibitem[Sugimori et al.(2009)]{SHS09}
Sugimori, T., Hyuga, H., Saito, Y.\yjourN{2009}{Phys. Soc. Jap.}{78}{034003}
{Fluctuation induced homochirality in an open system}

\bibitem[Sun et al.(2009)]{Sun09}
Sun, H. J., Saccomanno, V., Hedlund, B., \& McKay, C. P.\yab{2009}{9}{443}
{446}{Stereo-specific glucose consumption may be used to distinguish between chemical and biological reactivity on Mars: A preliminary test on Earth}

\bibitem[Toxvaerd(2013)]{Tox13}
Toxvaerd, S.\yoleb{2013}{43}{391}
{409}{The role of carbohydrates at the origin of homochirality in biosystems}

\bibitem[Toxvaerd(2014)]{Tox14}
Toxvaerd, S.\yjcpN{2014}{140}{044102}
{Discrete dynamics versus analytic dynamics}

\bibitem[Ulbricht(1975)]{Ulb75}
Ulbricht, T. L. V.\yol{1975}{6}{303}
{315}{The origin of optical asymmetry on Earth}

\bibitem[Valdivia-Silva(2012)]{Valdivia-Silva}
Valdivia-Silva, J. E., Navarro-Gonz\'alez, R., de la Rosa, J.  McKay, C. P.\yjour{2012}{Adv. Spa. Res.}{49}{821}
{833}{Decomposition of sodium formate and {\sc l}- and {\sc d}-alanine in the Pampas de La Joya soils: Implications as a new geochemical analogue to Martian regolith}

\bibitem[Voosen(2018)]{Voosen}
Voosen, P.\ysci{2018}{360}{1055}
{1055}{NASA Curiosity rover hits organic pay dirt on Mars: Carbon molecules in rocks from ancient lakebed resemble
kerogen, a ``goopy'' fossil fuel building block on Earth}

\bibitem[Walker(2017)]{Wal17}
Walker, S. I.\yrrpN{2017}{80}{092601}
{Origins of life: a problem for physics, a key issues review}

\bibitem[Zeldovich et al.(1977)]{ZSS77}
Zeldovich, B. Ya., Saakyan, D. B., \& Sobelman, I. I.\yjour{1977}{JETP Lett.}{25}{94}
{96}{Energy difference between right-hand and left-hand molecules, due to parity nonconservation in weak interactions of electrons with nuclei}

\end{thebibliography}
\end{document}